\documentclass[useAMS,usenatbib]{mnras}
\usepackage{times,graphicx,amsmath,amsfonts,amssymb,aas_macros,epstopdf}
\usepackage{epsfig}
\usepackage[usenames,dvipsnames]{xcolor}
\usepackage{url}
\usepackage{subfigure}
\usepackage[normalem]{ulem} 

\def\etal{{\frenchspacing\it et al.}}
\def\ie{{\frenchspacing\it i.e.}}
\def\eg{{\frenchspacing\it e.g.}}

\def\be{\begin{equation}}
\def\ee{\end{equation}}
\def\ba{\begin{eqnarray}}
\def\ea{\end{eqnarray}} 
\def\Msun{h^{-1}{\rm M}_{\odot}}
\def\hmpc{h^{-1}\,{\rm Mpc}}
\def\kms{\, {\rm km }\, {\rm s}^{-1}}

\def\hkpc{h^{-1}\,{\rm kpc}}

\def\lcdm{\Lambda{\rm CDM}}

\def\ln{{\rm ln}\,}

\def\frac#1#2{{\textstyle{#1\over #2}}}

\def\simlt{\stackrel{<}{{}_\sim}}
\def\simgt{\stackrel{>}{{}_\sim}}
\def\br{\mathbf{r}}
\def\bv{\mathbf{v}}
\def\bx{\mathbf{x}}
\def\bk{\mathbf{k}}
\def\rw{r_w}
\newcommand{\gadgetII}{\textsc{gadget2}}

\newcommand{\LGO}{\textsc{\bf LGO}}
\newcommand{\LGOI}{\textsc{\bf LGO1}}
\newcommand{\LGOII}{\textsc{\bf LGO2}}
\newcommand{\LGOIII}{\textsc{\bf LGO3}}
\newcommand{\LGOIV}{\textsc{\bf LGO4}}
\newcommand{\LGOV}{\textsc{\bf LGO5}}
\newcommand{\LGONOV}{\textsc{\bf LGO--NOV}}
\newcommand{\RNDO}{\textsc{\bf RNDO}}

\title[Velocity statistics for LG-like observers]{Not a Copernican observer:\\ biased peculiar velocity statistics in the local Universe}
\author[Wojciech A.~Hellwing \etal]{Wojciech A.~Hellwing$^{1,2}$\thanks{E-mail: pchela@icm.edu.pl}, Adi Nusser$^{3}$, 
Martin Feix$^{4}$ and Maciej Bilicki$^{5,2}$\\ 
$^{1}$Institute of Cosmology and Gravitation, University of Portsmouth, Portsmouth PO1 3FX, UK\\
$^{2}$Janusz Gil Institute of Astronomy, University of Zielona G\'ora, ul. Szafrana 2, 65-516 Zielona G\'ora, Poland\\
$^{3}$Department of Physics and the Asher Space Research Institute, Israel Institute of Technology Technion, Haifa 32000, Israel\\
$^{4}$CNRS, UMR 7095 \& UPMC, Institut d’Astrophysique de Paris, 98 bis Boulevard Arago, 75014 Paris, France\\
$^{5}$Sterrewacht Leiden, Universiteit Leiden, Niels Borhweg 2, NL-2333 CA Leiden, the Netherlands
}

\begin{document}
\date{Accepted XXXXXX. Received XXXXXX; in original form XXXXXX}
\pagerange{\pageref{firstpage}--\pageref{lastpage}} \pubyear{2016}
\maketitle
\label{firstpage}

\begin{abstract}
We assess the effect of the local large scale structure on the estimation of two-point 
statistics of the observed radial peculiar velocities of galaxies. 
A large N-body simulation is used to examine these 
statistics from the perspective of random observers as well as 
``Local Group (LG)-like" observers conditioned to reside in an environment
resembling the observed universe within 20 Mpc. 
The local environment systematically distorts the shape and amplitude of velocity statistics 
with respect to ensemble-averaged measurements made by a Copernican (random) observer.
The Virgo cluster has the most significant impact, introducing large systematic 
deviations in all the statistics. For a simple ''top-hat`` selection function,
an idealized survey extending to $\sim 160\hmpc$ or deeper is needed 
to completely mitigate the effects of the local environment.
Using shallower catalogues leads to systematic deviations 
of the order of $50$ to $200\%$ depending on the scale considered.
For a flat redshift distribution
similar to the one of the CosmicFlows-3 survey, the deviations are even 
more prominent in both the shape and amplitude 
at all separations considered $(\simlt 100\hmpc)$.
Conclusions based on statistics calculated without 
taking into account the impact of the local environment should be revisited.
\end{abstract}


\section{Introduction}
\label{sec:intro}
A pillar of cosmology is the Cosmological Principle \citep{Milne1935} stating that the Universe approaches 
isotropy and homogeneity with increasing scales\footnote{A counter example to the Cosmological Principle 
is a distribution of particles in a random fractal encompassing empty volumes of the
same size as the whole probed region \citep{1980Peebles,NusserLahav2000}}.
This principle is incorporated in the modern hierarchical scenario for structure formation, where matter 
density fluctuations are well defined, with a correlation function approaching zero on large scales. 
In such a scenario, initial fluctuations are described by homogeneous
Gaussian random fields, and thus measurements made by different random observers are equivalent. 
The difference in the statistical properties inferred by
these observers is commonly denoted as ``cosmic variance''. Assuming that our position 
in the Universe is not privileged, which is expressed in terms of the Copernican
Principle \citep[e.g.][]{Uzan2009}, deep large-scale galaxy redshift surveys
\citep[\eg][]{2dfgrs,sdss,Scrimgeour2012,CMASS_corr2012,Nadathur2013,Guzzo2014,Alpaslan2014} 
as well as detailed analyses of the cosmic microwave background radiation \citep[][]{WMAP9,Planck2013,Planck2015} 
broadly support this picture.
 
Recent years have witnessed the advent of high-quality and rich galaxy peculiar velocity data,
\eg{} the SFI++ \citep{Springob2007}, 6dF \citep{Springob2014},
and CosmicFlows catalogues \citep{Courtois2011,Tully2013,Tully2016}. This re-kindled activity 
in the peculiar velocity field with the new data offering an unprecedented opportunity
for cosmological measurements and theory testing. 
In late-time linear theory, peculiar velocities are proportional to the gravitational force field.
Therefore, peculiar velocity catalogues are a direct probe of dark matter 
and can in principle provide valuable information on fundamental theories 
for structure formation \cite{StraussWillick}. 

Inference of cosmological information from local observations must take into account the 
uncertainties introduced by cosmic variance. This has been known for
a long time, dating back to early studies of the density field of galaxies 
\citep[\eg][]{Sandage1978,Huchra1983,Soifer1984,Geller1989}. While 
cosmic variance in the statistical analysis of the galaxy distribution is well studied, its implications 
on peculiar velocity observations  have received little attention 
\cite[but see][]{Tormen1993,Strauss1998,Bilicki2010,Hellwing2014}
and remain poorly understood. Due to the long-range nature of gravity, 
local structures affect velocity correlations on much larger scales than 
those relevant to the density field \citep{Tormen1993,Borgani2000,Chodorowski2002}. 
With reliable velocity catalogues only available for galaxies out to distances 
of 100--200 $\hmpc$, the impact of nearby structures is likely very significant.
A similar effect was already hinted for the case of a local velocity
field dispersion measure \citep{Cooray2006,Marra2013,Wojtak2014}.

Galaxy peculiar velocities are practically unbiased 
with respect to the underlying velocity field 
\citep[\eg][]{Vittorio1986,Gorski1988,Groth1989,StraussWillick,Nusser1998,
streaming_vel_Omega,Juszkiewicz2000,Sarkar2007,Nusser2011,Hudson2012,Nusser2012,Feix2015}.
This is in contrast to the galaxy distribution in redshift surveys which is a biased tracer 
of the mass density field. Thus, peculiar velocity catalogues are 
not merely complementary to redshift-space distortions, but provide an independent
avenue towards testing fundamental 
physical theories of structure formation, dynamical dark energy and modified gravity 
\citep{infall_Zu2013,Li2013,Hellwing2014PhRvL,Berti2015,Bull2016}.

Extracting cosmological information from the observed motions is, however, a highly non-trivial matter. 
Despite the recent increase in quality and number of distance indicator measurements, the corresponding peculiar velocity
catalogues remain relatively sparse with significant observational and systematic errors especially at larger distances. 
There are several approaches for inferring cosmological information from the observations. One could make an attempt at 
reconstructing a 3D peculiar velocity field from which the underlying mass density can be derived. 
This would be very rewarding but the effort is hampered by the notorious inhomogeneous 
Malmquist bias \citep{Lynden-Bell1988a,Lynden-Bell1988b} leading to
spurious enhancement of the derived density fluctuations. 
A more straightforward strategy which has provided important constraints on the standard paradigm 
is to compare between the measured velocities and 
the gravitational field associated with an independent redshift survey \citep[see \eg][]{Davis2011}. 
Although this analysis is free from 
cosmic variance uncertainties and is mainly free from Malmquist biases, it relies on 
redshift surveys and is therefore dependent on the biasing relation between mass and galaxies.

Our main goal in this paper is to systematically assess the impact of cosmic variance 
and observer location on the peculiar velocity observables 
such as velocity correlation functions 
and mean streaming velocities (the first moment of galaxy pairwise velocity distribution).

We neglect meagre redshift evolution which might be present in local ($z\approx 0$) 
peculiar velocity catalogues. 
Further, we make no attempt at incorporating observational errors on the measured velocities.
These errors increase with distance and 
can obviously lead to large uncertainties. Subsequently, we do not model any 
inhomogeneous Malmquist bias related to these errors. 

This paper is organised as follows: in \S\ref{sec:simulation} we describe the numerical assets 
used in this work; section \S\ref{sec:vel_stats} introduces and describes velocity
statistics we consider; in \S\ref{sec:biases} we discuss various theoretical biases, while
in \S\ref{sec:LG_observers} we study the impact of observer location and 
galaxy radial selection on the velocity statistics.
We conclude with a general discussion of our results and their implications in \S\ref{sec:conclusions}.

\section{Simulations}
\label{sec:simulation}
Ideally we would like to study the velocity field of galaxies themselves. However, realistic modelling
of galaxy formation physics in a computer simulation is very difficult and computationally challenging.
Hence we will use here DM haloes and their peculiar velocities as proxies 
for luminous galaxies. In principle, such approach could hinder our analysis by introducing
systematic biases reflecting the fact that we ignore all the complicated baryonic physics.
Energetic feedback processes such as Active Galactic Nuclei,
star formation together with dynamical gas friction and ram pressure striping 
could significantly affect the velocities of visible (stellar) components of galaxies with respect to
their DM halo hosts. However, recently \citet{Hellwing2016} using EAGLE, the state-of-the-art
galaxy formation simulation \citep{Schaye2015}, have shown that peculiar velocities
of galaxies inhabiting haloes with $M_{200}>2\times 10^{11}\Msun$ are on average affected
by the baryonic effects at the level of at most 1 km/s, while even smaller (dimmer)
galaxies are affected at the level of at most a $10-20$ km/s. For all our practical purposes
such small effects would have negligible impact on our analysis,
indicating that we can safely ignore baryonic effects and model the galaxy peculiar
velocity field using DM haloes as their proxies.

We will base our analysis on a new $\lcdm$ N-body simulation
dubbed ``Warsaw Universe''. The detailed description of this resource will be presented
in an accompanying paper (Hellwing in prep.). Here we will limit ourselves to presenting only
the most important aspects of this simulation relevant for our study. The simulation
consists of 2 billion DM particles ($1280^3$) placed in a uniform cube of $800\hmpc$
width. It was evolved using publicly available \gadgetII{} code \citep{Gadget2}. The initial 
conditions were set at $z=63$ using the Zel'dovich approximation \citep{ZA}.
The initial density fluctuations power spectrum was chosen to follow 
WMAP7 best-fit values of cosmological parameters \citep[][data wmap7+bao+h0]{WMAP7}:
$\Omega_0h^2=0.134, \Omega_bh^2=0.0226, \Omega_{\lambda}=0.728, \sigma_8=0.809, n_s=0.963, h=0.704$.
In this work only the final snapshot of the simulation ($z=0$) will be considered, as
we are interested in the local galaxy velocity field. Thus, the resulting
resolutions of the simulation are: $m_p=1.84\times 10^{10}\Msun$ for the mass and
$\varepsilon=20\hkpc$ for the force.

DM haloes have been identified by means of the phase-space Friends-of-Friends ROCKSTAR halo finder,
kindly provided to the public by \cite{Behroozi2013}. 
For the $z=0$ simulation output, ROCKSTAR gave a little more than $\sim5.5\times 10^6$ 
bound DM haloes with a minimum of 
20 particles per halo (\ie with minimum $M_{200}=3.7\times10^{11}\Msun$).
Here we define the halo mass
as $M_{200}= 4/3\pi R_{200}^3200\times\rho_c$, where the radius $R_{200}$ is
the distance from a halo centre enclosing a sphere
with an average density of $200\rho_c$ where $\rho_c=3H^2/8\pi G$ is the critical density.
The bulk velocity of each halo is taken as the velocity vector of its centre-of-mass.
In the analysis of distance indicator catalogues, galaxies in groups and clusters are usually grouped together. 
To match that we have excised satellite subhaloes from our halo catalogue. 

\section{Velocity statistics}
\label{sec:vel_stats}
In this section we will describe two velocity statistics that are our primary focus
in this work. Namely the velocity correlation functions and moments of pairwise velocity
distribution function. In principle the cosmological information is encoded in the full 
three dimensional velocity field of galaxies. However, this is not accessible by astronomical 
observations, with a few exceptions in the very local Universe (Local Group)
\footnote{But see \citet{Nusser2012} for near-future prospects of measuring transverse velocities with Gaia.}.
Hence we need to limit ourselves to only the radial component of the peculiar
velocity field, which is a projection of the full 3D velocity vectors
onto the line of sight connecting an observer with an object in question.

We set the scale factor, $a$, to unity at the present time and denote 
the corresponding Hubble constant with 
$H_0$. The peculiar velocity of a test particle is 
$\dot{\bx}$ where $\bx$ is the comoving position of the particle. 
The density contrast is $\delta(\mathbf{x}) =\rho(\mathbf{x})/\bar \rho -1$ where $\rho(\mathbf{x})$ 
is the local density and $\bar \rho$ is the mean background density.

\subsection{Velocity Correlation Functions}
\label{subsec:vel_corr}
The correlation properties of a 3D peculiar velocity field, $\mathbf{v}(\mathbf{x})$, are specified
by the velocity correlation tensor
\be
\Psi_{ij}(\mathbf{r})\equiv\langle v_i(\mathbf{x}) v_j(\mathbf{x}+\mathbf{r})\rangle\,\,,
\label{eqn:vel_corr_tensor}
\ee
where $i,j$ are Cartesian components of $\bv$ and $\br$ is the separation between two points in space. 
For a statistically homogeneous and isotropic velocity field the velocity
correlation tensor can be written as a linear combination of
parallel (to the separation vector), $\Psi_\parallel$, and transverse, $\Psi_\perp$,
velocity correlation functions \citep[][]{Gorski1988}
\be
\Psi_{ij}(r)=\Psi_\perp(r)\delta_{ij}+\left[\Psi_\parallel(r)-\Psi_\perp(r)\right]\hat{r}_i\hat{r}_j\,\,,
\label{eqn:vel_cor_para_trans}
\ee
where $\delta_{ij}$ is the Kronecker delta.

In linear theory, the velocity correlations can easily be expressed in terms of 
the power spectrum $P(k)$ of the density fluctuations $\delta(\mathbf{x})$. 
Linear theory relates the Fourier components of peculiar velocity and 
density fluctuation fields by \citep[\eg][]{1980Peebles}
\be
\label{eqn:lin_fur_cont}
\bv(\bk)=-i H_0 f {\hat{\bk}\over k}\delta(\bk)\,\,,
\ee
where $f\equiv$d$\ln D_+(a)/$d$\ln a$ is the growth rate of density perturbations.
This yields \citep{Gorski1988}:
\be
\label{eqn:psi_per_lin}
\Psi_\perp(r)={H_0^2f^2\over 2\pi^2}\int P(k){j_1(kr)\over kr}\textrm{d}r\,,
\ee
and
\be
\label{eqn:psi_par_lin}
\Psi_\parallel(r)={H_0^2f^2\over 2\pi^2}\int P(k)\left[{j_0(kr)\over kr}-2{j_1(kr)\over kr}\right]\textrm{d}r\, ,
\ee
where 
\begin{equation}
j_0(y)=\dfrac{\sin y}{y} \quad {\rm and} \quad j_1(y)=\dfrac{\sin y}{y^2}-\dfrac{\cos y}{y} \; .
\end{equation}
Thus, in principle, measurements of $\Psi_\parallel$ and $\Psi_\perp$ should provide constraints on
a combination of the cosmological power spectrum and the growth rate, independent of galaxy biasing. 

\subsubsection{Correlations from radial velocities}
Observations provide access to the radial (line of sight) components of the galaxy peculiar velocities.
Hence the transverse and parallel correlation functions cannot be measured directly. 
\citet{Gorski1989} and \citet{Groth1989} proposed alternative velocity correlation statistics which could readily be computed from 
the observed radial components. Given a sample of $N$ galaxies with positions $\mathbf{r}_\alpha$ and radial 
peculiar velocities $u_\alpha=\mathbf{v}_\alpha\cdot \hat{\mathbf{r}}_\alpha$ ($\alpha=1\cdots N$),
let the separation vector between two galaxies be $\mathbf{r}=\mathbf{r}_\alpha -\mathbf{r}_\beta$, 
and the corresponding subtended angles are $\cos\theta_{\alpha\beta}=\hat{\mathbf{r}}_\alpha\cdot\hat{\mathbf{r}}_\beta$
and $\cos \theta_{\alpha}=\hat{\mathbf{r}}\cdot \hat{\mathbf{r}}_\alpha$. 
Then these statistics are defined as \citep{Gorski1989}
\be
\psi_1(r)={\sum_{\alpha,\beta}u_\alpha u_\beta \cos\theta_{\alpha\beta}\over \sum_{\alpha,\beta}\cos^2\theta_{\alpha\beta}}\,,
\label{eqn:psi1}
\ee
and
\be
\psi_2(r)={\sum_{\alpha,\beta}u_\alpha u_\beta \cos\theta_\alpha\cos\theta_\beta \over 
\sum_{\alpha,\beta}\cos\theta_{\alpha\beta}\cos\theta_\alpha \cos\theta_\beta}\,,
\label{eqn:psi2}
\ee
where the summation covers all galaxy pairs with separation $r<|\mathbf{r}_\alpha- \mathbf{r}_\beta|<r+\Delta r$.
The ensemble average of either of 
 $\psi_{1,2}(r)$ is a linear combination of $\Psi_\perp(r)$ and $\Psi_\parallel(r)$, 
\be
\label{eqn:psi_liner_comb}
\Psi_{1,2}(r)\equiv\langle\psi_{1,2}(r)\rangle=X_{1,2}(r)\Psi_\parallel(r)+\left[1-X_{1,2}(r)\right]\Psi_\perp(r)\,,
\ee
where the geometrical factors $X_{1,2}$ 
can be estimated directly from the data
\ba
X_{1}(r)={\sum_{\alpha,\beta}\left[\mathbf{r}_\alpha\mathbf{r}_\beta(\cos^2\theta_{\alpha\beta}-1)
+r^2\cos\theta_{\alpha\beta}\right]\cos\theta_{\alpha\beta}\over
r^2\sum_{\alpha,\beta}\cos^2\theta_{\alpha\beta}}\,,\\
\label{eqn:A_of_r}
X_{2}(r)={\sum_{\alpha,\beta}\left[\mathbf{r}_\alpha\mathbf{r}_\beta\left(\cos^2\theta_{\alpha\beta}-1\right) 
+ r^2\cos^\beta\theta_{\alpha\beta}\right]^2\over
r^2\sum_{\alpha,\beta}\left[\mathbf{r}_\alpha\mathbf{r}_2\left(\cos^2\theta_{\alpha\beta}-1\right) 
+ r^2\cos^2\theta_{\alpha\beta}\right]\cos\theta_{\alpha\beta}}\,,
\label{eqn:B_of_r}
\ea

The prescription for deriving the continuous limit of these expressions is to replace the summation over 
particles with integration over space as follows
\ba
\sum_\alpha \left(\cdots \right) \rightarrow \int d^3 \mathbf{r}_\alpha n_{\rm obs}(\mathbf{r}_\alpha) \left(\cdots \right)\; .
\label{eqn:contlim}
\ea
Here, $n_{\rm obs}=\bar n (1+\delta_{\rm g}) \phi$ is the observed number density of galaxies and it is the product of the 
underlying number density $\bar n (1+\delta_{\rm g}) $ and the selection function imposed on 
the observations, $\phi$. Since galaxies are biased tracers of mass, 
the contrast $\delta_{\rm g}$ differs from the mass density contrast $\delta$.
Therefore, although the expressions (\ref{eqn:psi1}) and (\ref{eqn:psi2}) for $\psi_{1,2}$ are 
straightforward to compute from 
a velocity catalogue, the task of inferring cosmological information is quite challenging and difficult. 

\subsection{Pairwise velocity correlation}
\label{subsec:pairwise}
The other velocity statistics that we consider is the first moment
of the galaxy/halo pairwise velocity distribution. It is sometimes dubbed
as \textit{pairwise streaming velocity} and indicated as $v_{12}$. 
This statistic was introduced by \cite{DavisPeebles_BBGKY} 
in the context of the {\it Bogoliubov-Born-Green-Kirkwood-Yvon hierarchy} (BBGKY),
a kinetic theory which describes the dynamical evolution of a
system of particles interacting through gravity. 
This statistic is of special importance for modelling the correlation function of galaxies in 
redshift space. Here we will 
focus on its use as a characteristic of the flow pattern as probed by observed radial motions. 
We begin with the definition of this function in the fluid limit where we are given 
the full velocity and density fields.
In this idealized situation we write 
\be
\label{eqn:v12-weighted}
\mathbf{v}_{12}(r) = \langle\mathbf{v}_1-\mathbf{v}_2\rangle_{\rho} =
{\langle(\mathbf{v}_1-\mathbf{v}_2)(1+\delta_1)(1+\delta_2)\rangle\over
 1+\xi(r)}\,\,, 
\ee 
where $\mathbf{v}_1$ and $\delta_1=\rho_1/\langle\rho\rangle-1$ denote
the peculiar velocity and fractional matter density contrast 
at galaxy/halo position $\mathbf{r}_1$. Further 
$
\xi(r)=\langle\delta_1\delta_2\rangle
\label{eqn:xi2}
$
is the usual 2-point density correlation function. 
The $\langle\cdots\rangle_{\rho}$ denotes a 
pair-weighted average, which differs from the usual spatial averaging by
the weighting factor, 
$\mathcal{W}=\rho_1\rho_2/\langle\rho_1\rho_2\rangle$, which 
is proportional to the number density of pairs. 
Isotropy implies that $\mathbf{v}_{12}$ has a vanishing component in the perpendicular 
direction to the separation $\mathbf{r}$, i.e. 
$\mathbf{v}_{12}=v_{12}\hat{\mathbf{r}}$

In the stable clustering regime, on scales where the pairwise
velocity exactly cancels out the Hubble flow, $v_{12}=-H\mathbf{r}$.
The pair conservation equation \citep{1980Peebles} connects $v_{12}(r)$ 
to the density correlation function $\xi(r)$.
\citet{Juszkiewicz1999} suggested an analytical {\it ansatz} for Eqn.~(\ref{eqn:v12-weighted}),
which turned out to be a reasonably good approximation to results from N-body simulations 
 evolved from initial Gaussian conditions. Their formula reads
\be
v_{12}=-{2\over3}H_0rf\bar{\bar{\xi}}(r)[1+\alpha\bar{\bar{\xi}}(r)],
\label{eqn:juszkiewicz_anstatz}
\ee
where
\be
\label{eqn:aver_xi2}
\bar{\xi}(r)=(3/r^3)\int_0^r\xi(x)x^2dx\equiv\bar{\bar{\xi}}(r)[1+\xi(r)]\,\,.
\ee
Here $\alpha$ is a parameter that depends on the logarithmic slope of
$\xi(r)$. 
It is clear that $v_{12}(r)$ is a strong function of $\xi(r)$ and $f$.
Because of this some authors have suggested to use $v_{12}(r)$ as a cosmological probe
\citep{Feldman2003,Juszkiewicz2000,Hellwing2014PhRvL,Ma2015,Ivarsen2016}

\subsubsection{Pairwise correlation from radial velocities}
Using a simple least-square approach, \citet{Ferreira1999} derived an
estimator of the mean pairwise velocity applicable to catalogues of observed
radial peculiar velocities. It takes the following form:
\be
\label{eqn:v12_obs_estim}
\tilde{v}_{12}(r)={2\sum_{\alpha,\beta}(u_\alpha-u_\beta)p_{\alpha\beta}\over\sum_{\alpha,\beta}p^2_{\alpha\beta}}\,.
\ee
Here $p_{\alpha\beta}\equiv\hat{r}\cdot(\hat{\mathbf{r}}_\alpha+\hat{\mathbf{r}}_\beta)=
\cos\theta_\alpha+\cos\theta_\beta$. The continuous limit of the expression (\ref{eqn:v12_obs_estim}) 
is obtained from the recipe in (\ref{eqn:contlim}). 
Therefore, like $\psi_{1,2}$, this estimator depends on the underlying galaxy distribution
as well as the selection criteria.

\section{Estimator biases for randomly selected observers}
\label{sec:biases}
\begin{figure}
 \includegraphics[angle=-90,width=85mm]{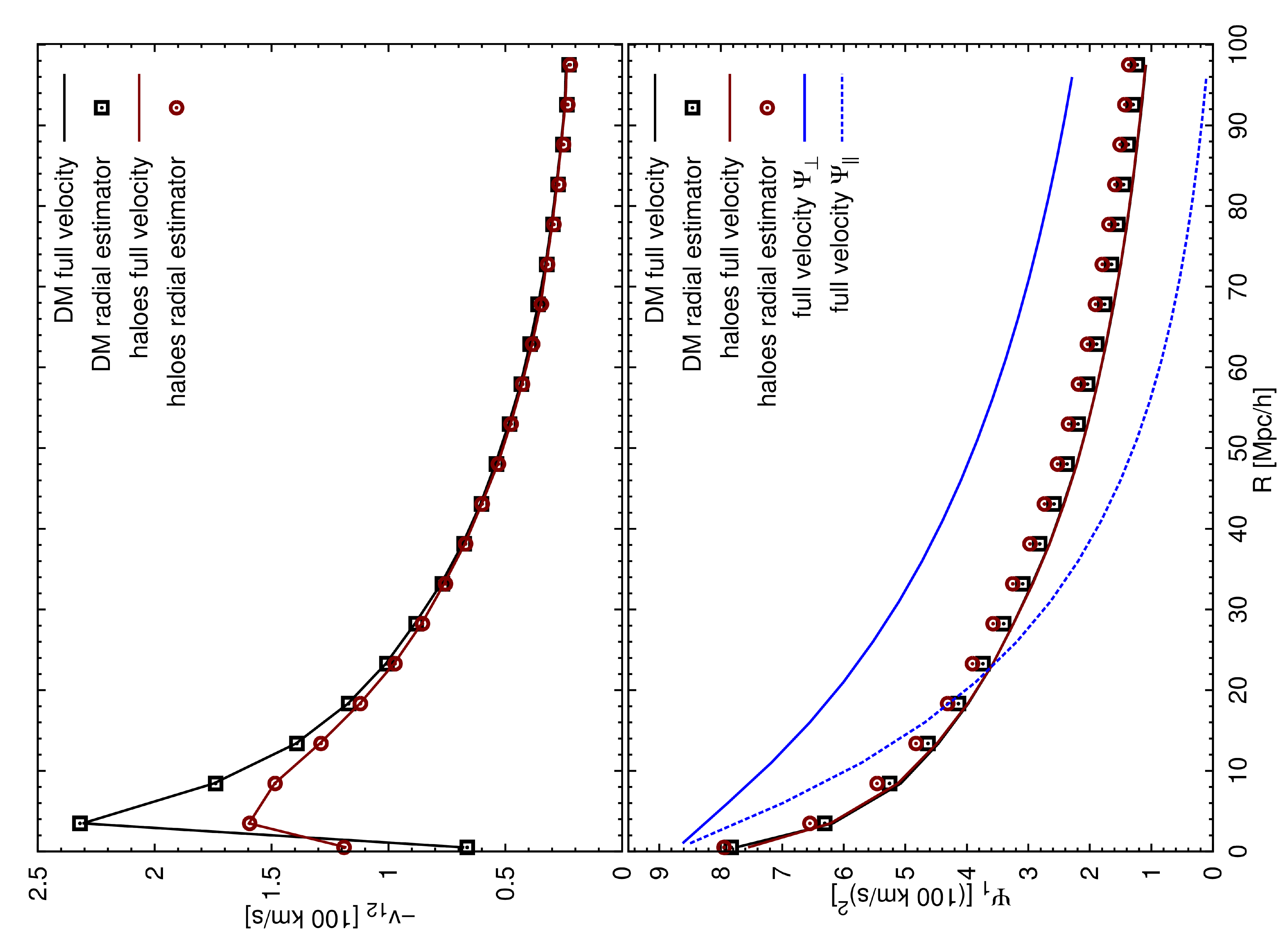}
 \caption{ 
 The performance of the radial velocity-based estimators compared 
 to statistics extracted using full velocity information.
 The top panel is for the pairwise velocities $v_{12}$, and the bottom for the velocity
 correlation function $\psi_1$. In all cases the lines depict 
 results obtained from full 3D velocity information. 
 Open squares and circles correspond to the radial-component 
 based estimators of Eqn.~(\ref{eqn:v12_obs_estim}) and Eqn.~(\ref{eqn:psi1})
 for DM and halo velocities, respectively.}
\label{fig:estimators_test}
\end{figure}
\begin{figure}
 \includegraphics[angle=-90,width=85mm]{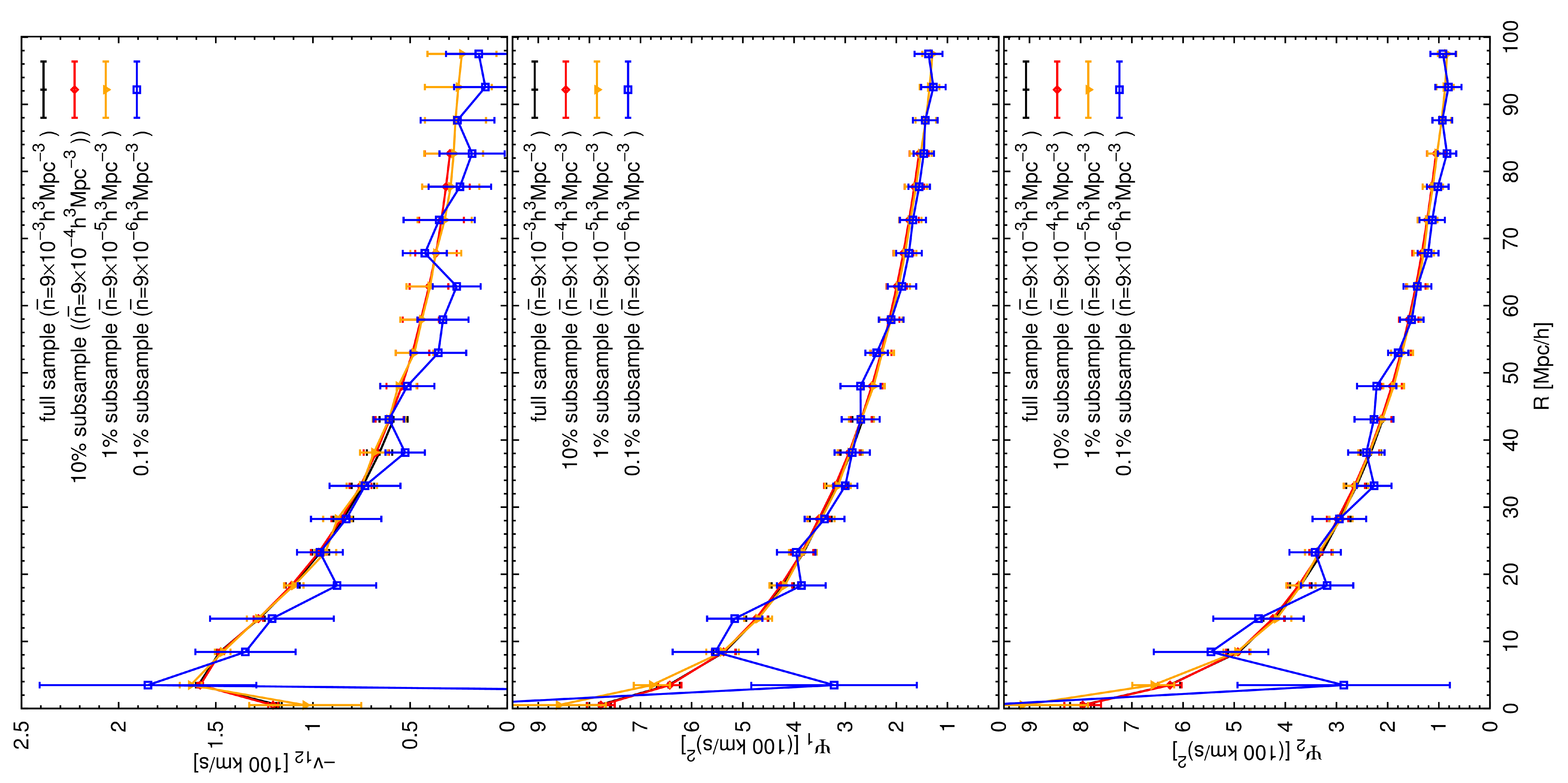}
 \caption{Sampling bias for $v_{12}(R)$ (top panel), $\Psi_1(R)$ 
 (middle panel) and $\Psi_2(R)$ (bottom panel).
   Different lines correspond to different halo samples.}
\label{fig:sampling_bias1}
\end{figure}
We begin our analysis be assessing how accurately the radial velocity based estimators
probe the true underlying 3D quantities. 
We consider 50 observers randomly placed in the simulation box of $800\hmpc$. We use 
the full halo catalogue with a minimum halo mass of $3.7\times10^{11}\Msun$, and compute the halo radial velocities 
relative to each observer. 
Because the radial velocity is observer-dependent, the radial velocity correlations
are expected to depend on the location of the observer. 
We compute the {\it ensemble} average over all the 50 observers. 
We treat such an averaged measurement as one made by
{\it the idealized Copernican observer}.
This ensemble average is then compared with the correlation function obtained 
from the full 3D velocity data of the full halo catalogue. 

The results are shown in Fig.~\ref{fig:estimators_test}.
In the top panel the radial component 
based estimator of (\ref{eqn:v12_obs_estim}) for $v_{12}$ is
shown (open symbols) against the result
(solid lines) obtained 
by summing over the same pairs in the simulation but using the full 3D velocity information.
We present separately results for DM particles (squares) and haloes (circles), 
as indicated in the panel. The agreement between the 
radial velocity and theoretical estimators is superb. 
For tracers, DM and haloes, and on all considered pair separations up to $100\hmpc$,
the differences between the radial component estimator for $v_{12}$ and 
the values obtained using full 3D information are smaller than $1-2$ km/s.
The bottom panel illustrates analogous comparison for $\psi_1$. Because
the results for $\psi_2$ follow quantitatively those of $\psi_1$,
we omit them for clarity. Since $\psi_1(r)$ is by construction defined
only for radial velocities, to get a theoretical prediction to compare with
we use Eqn.~(\ref{eqn:psi_liner_comb}) and Eqn.~(\ref{eqn:A_of_r}).
Here we computed $\Psi_\perp$ and $\Psi_\parallel$ directly from
the full 3D velocity field and used them together with the measured
geometrical factor $X_1$ to obtain a prediction
for $\psi_1$ (which we mark as 'full velocity' lines).
Unlike the previous case, the estimators for
the velocity correlation functions are slightly biased towards
higher values. Although noticeable, the effect is not large.
For DM particles it is less than $4\%$ at $R<20\hmpc$,
increasing to $\sim 8\%$ at $60\hmpc$. For haloes, the discrepancy
is roughly twice as large. Hence at scales of $60\hmpc$, it can be of the 
order of $15\%$, which should be taken into account, when one
wants to compare $\Psi_\perp$ and $\Psi_\parallel$ derived 
from measured $\psi_{1,2}$ with theoretical predictions 
of Eqn.~(\ref{eqn:psi_per_lin}) and Eqn.~(\ref{eqn:psi_par_lin}).

Having checked that both our radial velocity based estimators preform
reasonably well using the full halo catalogue, we now examine effects of the sparse halo sampling. 
Modern galaxy redshift surveys already contain millions of galaxies,
however such a sampling rate is far from the reach of velocity catalogues,
consisting of only thousands of objects. 
Nevertheless, despite the much lower object counts the velocity catalogues
retain quite high number density of tracers thanks to relatively small and limited volumes
that they cover. The currently available velocity catalogues are typically
reaching $\bar{n}\approx 10^{-4}-10^{-5}\,{\rm Mpc}^{-3}$.
However, such catalogues often need to be further diluted, when one needs
to for example reject galaxies with large velocity errors.
To assess how our velocity statistics and their estimators are 
affected by sub-sampling we split our full halo catalogue
into three randomly sub-sampled populations. In all the cases, we use
the original catalogue and sub-samples 
containing respectively 10\%, 1\%, and 0.1\% of the full sample.
The corresponding spatial abundances of resulting catalogues are:
$\bar{n}_{full,10\%,1\%,0.1\%}=9\times(10^{-3},10^{-4},10^{-5},10^{-6})h^{3}\,{\rm Mpc}^{-3}$, 
respectively. The three panels of Fig.~\ref{fig:sampling_bias1} 
illustrate the effect of sparse sampling,
from top to bottom for $v_{12}$, $\psi_1$ and $\psi_2$. 
As previously stated, all the plotted lines are ensemble averages
over 50 random observers, with the error bars marking $1\sigma$
dispersions around the ensemble mean. Analysis of the data
shown in plots reveals that the sub-sampling
only increases the scatter, while averages of both $v_{12}$ and $\psi_{1,2}$ are
not affected in any significant way. Only for the case of
the most diluted sample with only $1/1000$-th of the original haloes
appreciable scatter around the true mean of $v_{12}$ appears.
The same sub-sample traces the averages of $\psi_1$ and $\psi_2$ much better,
 already at $R\geq20\hmpc$ the effects of sparse sampling 
are small. Furthermore, it is noteworthy that the additional scatter due to sparse sampling is only
prominent for small separation bins, indicating that this scatter 
is sub-dominant to the cosmic variance. 
Hence we can safely expect that for $R\simgt 10\hmpc$ the velocity correlation functions are
well probed even with samples hundred times scarcer 
than the complete volume selection sample. 
These are good news, as we can now expect that relative
low sampling rate in the galaxy peculiar velocity surveys should not
affect significantly the measured velocity correlations. 

\section{Local Group Observers}
So far we have considered random observes in the box.
Now we turn to the effects of the nearby large-scale structure on the inferred velocity statistics. 
We, therefore, aim at selecting LG-analogue observers residing in regions resembling in as much as possible 
our local environment. 
The LG is a gravitationally bound system of a dozen major galaxies with the Milky Way (MW) 
and its neighbouring M31 as the most massive members.
The region of 5 Mpc distance from the LG 
 is characterized by moderate density \citep[see \eg][]{Tully1987,Tully1988,Hudson1993,Tully2008,Courtois2013}
and a quiet flow \citep{Sandage1972,Schlegel1994,Karachentsev2002,Karachentsev2003}. 
Located at a distance of $\sim 17$ Mpc is 
the Virgo cluster,
whose gravitational effects extend to tens of Mpcs around us, as evident from the corresponding 
infall flow pattern of galaxies \citep{TullyShaya1984,Tammann1985,Lu1994,Gudehus1995,Karachentsev2014}.
The presence of such a large non-linear mass aggregation
can have a substantial impact on peculiar velocity
field of the local galaxies.

To find suitable ``observers" in the simulation box we first 
obtain density and velocity fields on a regular $512^3$
grid by using the publicly available DTFE code \citep{cv2011}.
The DTFE code employs {\it the Delaunay Tessellation Field Estimation},
a method described in detail in \citet{sv2000,vs2009}, which assures that the resulting smooth fields 
have the highest attainable resolution, are volume weighted and have suppressed sampling noise.
The fields are then smoothed using top-hat filtering
and the resulting grid cells are used for imposing the local 
density and velocity constraints.
Given the density and velocity fields as well as the halo catalogue we search the simulation for candidate observers.
 Specifically we demand that ``observers" are located in an environment satisfying the following constraints:
\begin{enumerate}
\item \label{item:MW} the observer is located in a MW-like host halo of mass $7\times10^{11}<M_{200}/(\Msun)<2\times10^{12}$ 
\citep{Busha2011,Phelps2013,Cautun2014,GuoCooper2015},
 \item \label{item:v} the bulk velocity within a sphere of $R=3.125\hmpc$ centred on the observer is $V= 622 \pm 150\kms$ 
 \citep{Kogut1993},
 \item \label{item:d} the mean density contrast within the same sphere is in the range of $-0.2\leq\delta\leq 3$ 
 \citep{Karachentsev2012,Elyiv2013,Laniakea},
 \item \label{item:virg} a Virgo-like cluster of mass $M=(1.2\pm 0.6)\times10^{15}\Msun$ is present 
 at a distance $D=12\pm4\hmpc$ from the observer \citep{Tammann1985,Mei2007}.
\end{enumerate}
To examine the role of individual criteria we also study results for sets of observers selected 
 without imposing all constraints. The sets of observers we consider are: 
\begin{description}
\item[\LGOI] is our fiduciary set of 290 observers each satisfying all the selection 
criteria \ref{item:MW} through \ref{item:virg}. 
\item[\LGOII] consists of 1045 candidate observers obtained by relaxing the velocity constraint \ref{item:v}, 
but satisfying the remaining criteria. 
 \item[\LGOIII] has 804 candidates obtained by relaxing the density contrast condition 
 \ref{item:d} only. 
\item[\LGOIV] of 1561 candidates with the conditions \ref{item:v} \& \ref{item:d}
relaxed simultaneously. 
\item[\LGOV] has 1197 observers without imposing the constraint on the host halo 
mass but with all the other criteria fulfilled. 
\item[\LGONOV] contains 772543 candidate observers satisfying all conditions except the 
proximity to a Virgo-like cluster. 
\item[\RNDO] is a list of observers with randomly selected 
positions in the simulation box. This set is used as a benchmark for comparison. 
\end{description} 

Based on the number of candidate observers in each set, we conclude that 
the proximity to a Virgo-like cluster is the strongest discriminator among
all the conditions.
Moreover, positions of observers in each of the 5 sets {\LGOI}-{\LGOV} 
are highly correlated, as they are constrained to reside 
in the same vicinities of Virgo like objects. 
Therefore, in order to speed up the calculations, we consider only a 
sub-sample of the list of observes, not reducing however the statistical significance of the results. 
 This is done by laying a uniform coarse $8^3$ grid in the box and selecting, 
for each set of observers, one random observer per grid cell, should the cell 
contain any observers. This 
gives an average number of $60$ observers for each of the 5 sets. To match
the sample variance we also keep only 64 observers in the
\LGONOV{} and \RNDO{} sets. 
As we have already pointed out, currently available 
peculiar velocity catalogues are relatively shallow due to 
the difficulty in measuring distances especially for distant galaxies. 
Furthermore, additional distance cuts and trimming of the data 
are usually imposed on velocity catalogues in order to avoid 
very large errors and uncontrolled observational systematics.
To get closer to a realistic catalogue, 
we implement two simple data weighting schemes.
The first scheme mimics simple radial selection cuts that one
can always implement for a given peculiar velocity catalogue.
It is defined by a single ``depth" parameter, $\rw$.
Here a halo at a distance $r$ from the observer is assigned a weight, $w_h$, given by
\be
\label{eqn:data_weighting}
w_{h}=
\begin{cases}
1, & \text{if}\ r\leq\rw \\
0, & \text{otherwise}\, .
\end{cases}
\ee
The second scheme aims at mimicking a sample with a flattened radial distribution
of galaxies, similar to the one describing the CosmicFlows-3 catalogue \citep{Tully2016}.
Here, the weighting is characterised by a power-law and, in addition to the depth parameter $r_W$,
is also a function of the ''steepness`` parameter $m$. The corresponding formula
for $w_h$ is
\be
\label{eqn:data_weighting2}
w_{h}=
\begin{cases}
1, & \text{if}\ r\leq\rw \\
(r/\rw)^{-m}, & \text{otherwise}\, .
\end{cases}
\ee
Here we consider $\rw=20\hmpc$ and $m=2,3$ and dub the corresponding
catalogues {\it CF3-like m=2} and {\it CF3-like m=3} accordingly.
We will use these data weighting schemes to further investigate
how the velocity statistics depend on the catalogue depth.

\label{sec:LG_observers}
\begin{figure*}
 \includegraphics[angle=-90,width=170mm]{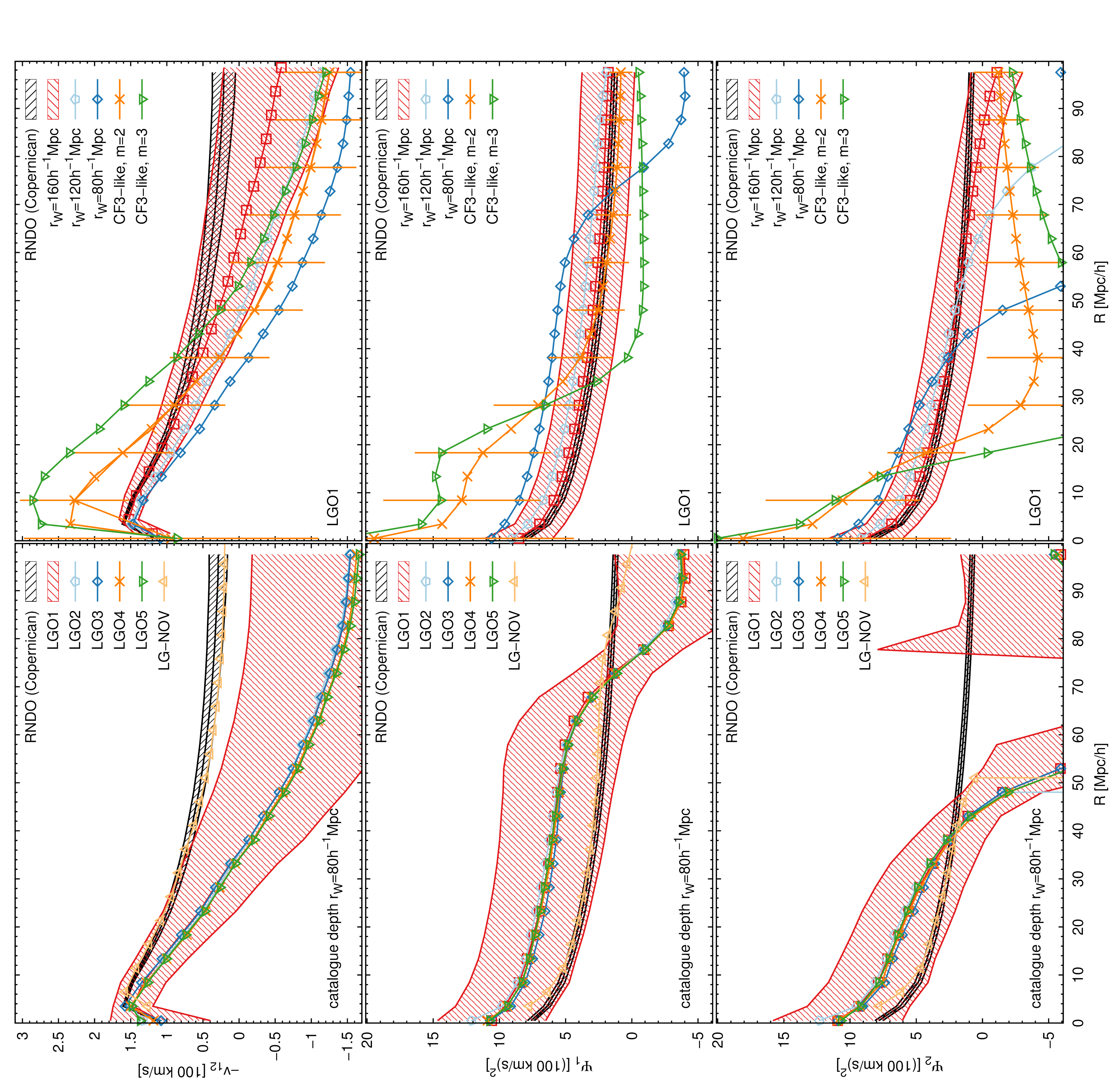}
 \caption{The effects of \LGO{}-like observer location and various selection functions on
 the velocity statistics. Error-bars and filled regions mark $1\sigma$ observer-to-observer scatter
 around ensemble mean. From to to bottom panels show results for $v_{12}$, $\psi_1$ and $\psi_2$ respectively.
 The left column illustrate the effects for different set of observers, but with the same imposed
 radial selection cut of $\rw=80\hmpc$. The right column of panels focuses on our main Local Group (\LGOI{})
 observers sample and the comparison of various selection functions and data weights.}
\label{fig:observers_bias}
\end{figure*}
\begin{figure}
 \includegraphics[angle=-90,width=85mm]{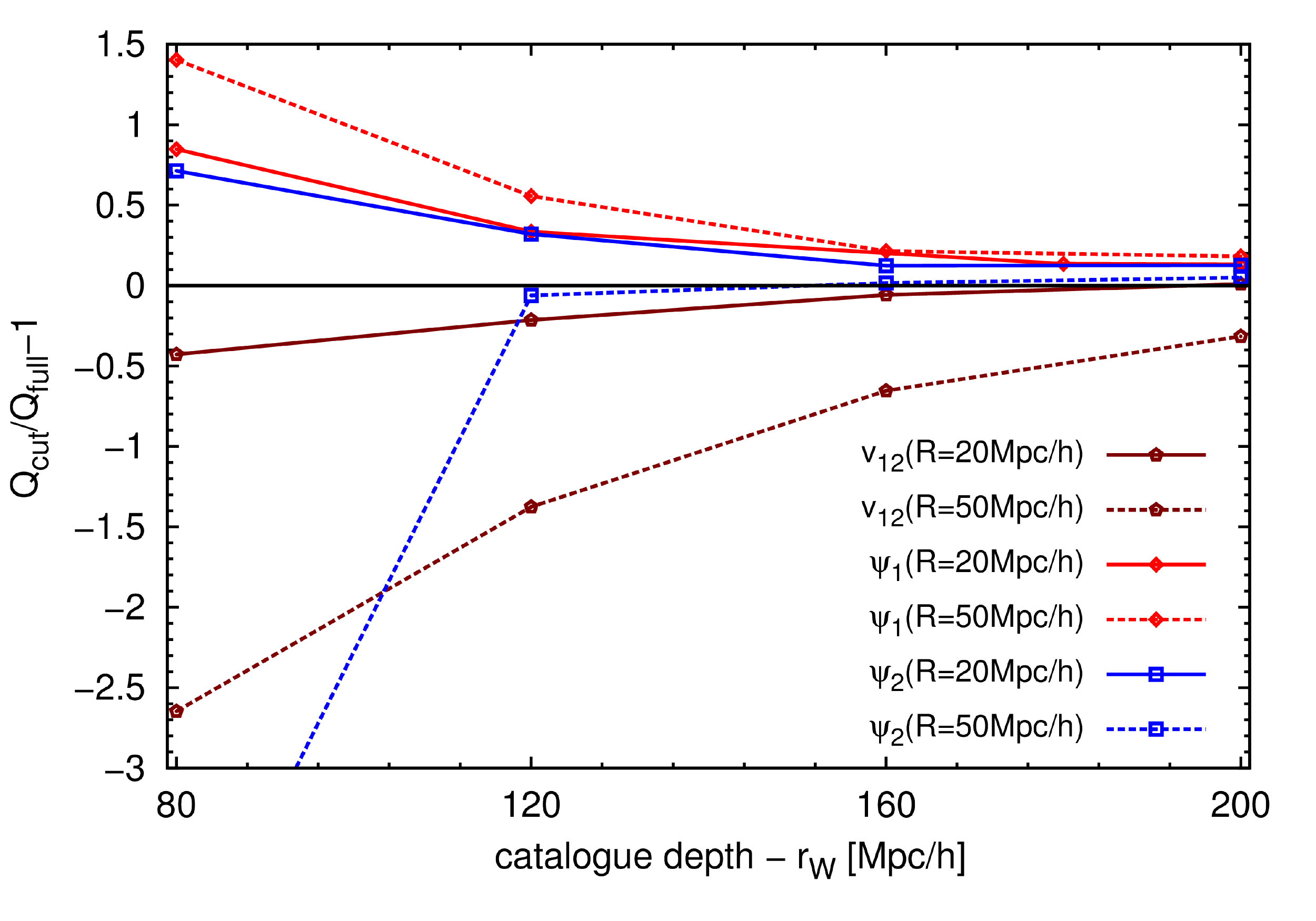}
 \caption{The biasing of the LG observer estimators with respect to a Copernican observer 
 as a function of the velocity catalogue depth parameter, $\rw$.
 The ratios for each estimator are taken at two pair separation scales: $20\hmpc$ and $50\hmpc$.}
\label{fig:rw_bias}
\end{figure}
 
The three panels to the left in Fig.~\ref{fig:observers_bias} show the statistics derived for 
all sets of observers, with $\rw=80\hmpc$ and the first weighting scheme applied. 
The curves are (ensemble) averages over all observers in each set 
(as indicated in the figure) and the attached error-bars and filled regions represent 
the corresponding $1\sigma$ scatter. 
The error-bars in the \LGO{} series are similar and for clarity they are attached only to \LGOI. 
Since we do not include observational errors, this scatter is entirely due to 
cosmic variance among the observers in each set. 
Plotted also are results for the ``Copernican'' observer,
computed from the full catalogue for the \RNDO{} observer set. 
The small error-bars here 
reflect the fact that different observers see different (radial) velocity components of the same galaxies.
The \LGO{} curves in all the panels differ systematically from the Copernican \RNDO{} result. However, the
\LGONOV{} and \RNDO{} curves are almost indistinguishable up to pair separations of $R\sim55\hmpc$,
meaning that the proximity to Virgo is the only significant criterion
in the selection of the LG candidate.
The average streaming velocity, $v_{12}$, defined in Eqn.~(\ref{eqn:v12_obs_estim}), 
in the top-left panel is significantly affected
by the LG selection criteria at pair separations $R\simgt 40\hmpc$. 
At those scales, LG-observers are deviating 
from the ``Copernican'' curve by more than $1\sigma$ getting values
lower then the cosmic mean observer.
However, the observer-to-observer induced variance is large. So even for
smaller scales, where the both averages agree within the scatter, the amplitude
of the difference is large and can typically take from 50 to 100 km/s.
This is already a $100\%$ level effect
at $R=40\hmpc$, but it quickly grows, reaching
$200\%$ magnitude difference already at separations of $\sim 60\hmpc$.
At large separations closer to $\rw$, fewer galaxy/halo pairs
are found which explains the rapid increase of the error-bars for LG-observers.

In the middle and bottom-left panels of Fig.~\ref{fig:observers_bias} we consider the correlation 
functions, $\psi_1$ and $\psi_2$.
For all \LGO{} LG-analogues the amplitude of $\psi_1$
is systematically larger than the black curve corresponding to 
the Copernican observer, up-to separations of $\sim75\hmpc$.
At larger separation the sign of the effect is flipped and all \LGO{} $\psi_1$'s
take smaller amplitudes than a random observer measurement.
This is a clear sign of the imposed catalogue depth, with our radial cut of $\rw=80\hmpc$. 
Here again the observer induced scatter is large making the \LGO{}
curves to ``agree'' within $1\sigma$ with the Copernican observer, even though 
the actual relative difference is typically as large as $\sim50\%$.
However, considering just the small variance of \RNDO, the \LGO{}
results would be $>5\sigma$ away from a cosmic mean.
For $\psi_2$ the behaviour is qualitatively similar to the $\psi_1$
case. The main difference consists of a roughly twice smaller scale ($\sim40\hmpc$)
at which the flip of the effect's sign occurs. 
However, the noteworthy feature of $\psi_2$ \LGO{} signal 
is the significantly smaller relative difference from \RNDO{}, which
typically takes only $25\%$ and also a slightly smaller observer-based variance.
Interestingly it seems that also ``no Virgo'' observers
for both $\psi$'s at scales above the ``flip off''
differ in the same way from the random observers results as \LGO{}-ones.
As we have already noticed, for all three estimators the scatter connected 
with a LG-like observer is much larger than for the random 
observer sample. We have checked that sampling variance is
not contributing significantly to this scatter, as all estimates
are based on comparable pair-number counts per bin.
This implies that even for the signal
extracted at large galaxy pair separations, the variance
induced by the local structures is large and significant.
This is an intrinsic LG-like observer property and
as such for a realistic case of one LG-observer,
this large scatter will manifest {\it as a systematic error}
on the velocity correlation functions.

The column to the right of Fig. 3
shows the same statistics obtained for one and the same main
\LGOI{} list, but with both data weighting schemes considered.
For the simplistic scheme of Eqn. (\ref{eqn:data_weighting}) 
we implement the following catalogue depths:
$\rw=80,120,160\hmpc$. In addition we also consider two
CF3-like samples with $m=2$ and $m=3$.
The right column of panels in Fig.~\ref{fig:observers_bias} shows
how the effective radial depth and related incompleteness affects $v_{12},\psi_1$ and $\psi_2$.
The behaviour of curves corresponding to different radial selection cuts
is qualitatively similar for all three panels. As expected,
the shallower the catalogue, the bigger the effect of observer location.
For CF3-like selection functions the effects of the observer's
location become more severe for $R\simlt 40\hmpc$, where both the scatter
and the relative differences are bigger than for the shallowest $\rw=80\hmpc$
\LGO{}-case. Yet, at larger pair separations it seems that the situation
is partially remedied, where (especially for $m=2$) the data from more distant
galaxies bring the curves again closer to \RNDO{}.
In contrast to the situation we have encountered for a simple $\rw=80\hmpc$
cuts presented in the left panels, where $\psi_2$ appeared as the least
affected statistics, here for a CF3-like selection function,
it is $\psi_1$ that is characterised by least biased behaviour.
For $R\geq40\hmpc$ its average is even consistent within $1\sigma$ 
with the Copernican observer's one.
Finally, as one might expect the difference between
random observers and the deepest $\rw=160\hmpc$ \LGOI{} catalogue 
are very small (when compared to differences visible for shallower catalogues).

To allow for a better assessment of the effect on 
the velocity statistics inferred by \LGOI{} observers,
we plot in Fig.~\ref{fig:rw_bias} the ratio of LG-based estimators with respect to Copernican
observers as a function of the catalogue depth parameter
$\rw$. We focus on ratios taken at two pair separations of $20$ and $50\hmpc$.
We also add a catalogue with $\rw=200\hmpc$, which benchmarks the limiting
case of an idealized very deep velocity survey.
It is clear now that both $\rw=80\hmpc$ and $120\hmpc$ samples
are dramatically affected by the limited depth of their halo catalogues.
For $80\hmpc$ depth catalogue the differences from the full depth one
can be typically as large as $>4\sigma$, while for $120\hmpc$ catalogue
the deviations from the unbiased case are contained in the range 
of $2-2.5\sigma$. The situation is better for the two deepest catalogues
we consider with cuts at $160$ and $200\hmpc$. However, here,
even for $200\hmpc$ case the differences between values inferred
from a realistic catalogue and an ``idealized'' deep one 
are bigger than $1\sigma$ for $R\simgt 50\hmpc$ in $v_{12}$ 
case, and $R\simgt 20\hmpc$ for $\psi_1$. 
For $\rw\geq120\hmpc$ and $R\geq50\hmpc$
the results for the $\psi_2$-estimator
seems to be the closest one to universal cosmic mean of \RNDO{}. We caution however,
that as indicated by the results shown in the bottom-right panel
of Fig.~\ref{fig:observers_bias} for a more realistic CF3-like selection
functions, $\psi_2$ at those large separations is more affected
than $\psi_1$.
In all cases the scatter due to observer location induced by limited depth of catalogues 
is large, and as expected grows with shrinking catalogue depth.

\section{Discussion and Conclusions}
\label{sec:conclusions}
In this paper we considered the estimation of two-point peculiar velocity statistics. 
We have refrained from assessing important effects related to observational errors 
such as Malmquist biases, and focused on the impact of cosmic variance and observer location. 

We have tested the ability of the radial velocity based estimators in
Eqns.~(\ref{eqn:psi1}), (\ref{eqn:psi2}) and (\ref{eqn:v12_obs_estim}) at recovering 
the underlying correlations in the case of 
of complete coverage velocity catalogues.
The $v_{12}$ estimator of \citet{Ferreira1999}
performs very well by measuring the averaged infall velocity with a percent-level accuracy.
The theoretical predictions for both correlations functions were off by a factor of
$8-16\%$. 
Thus, even for perfect data the measured values of $\psi_1$ and $\psi_2$ should be
compared with theoretical predictions of Eqn.~(\ref{eqn:psi_liner_comb}) with care.
Further, since for realistic data these statistics depend strongly on the data completeness, 
a much better approach is to derive predictions for both
\citet{Gorski1989} functions based on realistic mock catalogues, rather than a simplistic
relation as the one expressed 
by Eqns.~(\ref{eqn:psi_per_lin}), (\ref{eqn:psi_par_lin}) and (\ref{eqn:psi_liner_comb}).

Next we have checked if a sampling bias due to strong under-sampling would be an issue.
This was a relevant test, as the currently available galaxy peculiar velocity catalogues
contain a relatively small number ($\sim 10^4$) of objects.
The tests show that all three velocity statistics are not sensitive to
under-sampling.
The ensemble averages of 10\% and $1\%$-sub-samples 
(with effective $\bar{n}=9\times10^{-4}$ and $9\times10^{-5}\,h^3 {\rm Mpc}^{-3}$ number densities)
were statistically 
consistent with the full sample. Only in the case of a severe sub-sampling
of the $0.1\%$-case (with $\bar{n}=9\times10^{-6}\,h^3{\rm Mpc}^{-3}$)
the estimated mean showed some noticeable scatter around
the true mean.
In addition, we have found that the scatter around the mean
is scale dependent, being a strong function of a pair separation for $v_{12}$.
Albeit for both $\psi$'s, except the smallest scales of $R<30\hmpc$, the scatter shows
only a very weak evolution with scale. All in all, we can report that all the three
studied velocity statistics are performing well in the sparse sampling regime.

Our most important result is related to the effect of the observed large scale environment 
on velocity statistics. 
We have performed a detailed analysis of cosmic variance in velocity statistics 
by considering differences in velocity observables as measured by a
Copernican observer and LG-equivalents. We have considered four criteria compatible with 
LG properties and local environment. Velocity two-point
statistics are found to be insensitive to the criteria related to the MW halo mass and 
the LG motion and its mean density (within $\sim 3\hmpc$). In contrast, the proximity 
of an observer to a Virgo-like cluster is highly significant, affecting the correlations up to scales of $\sim100\hmpc$.
This has not been noticed by \citet{Tormen1993} since they only consider LG-analogue 
observers defined without imposing the presence of a nearby massive cluster. 

In the near future, peculiar velocity surveys are not likely 
to reach to much larger distances than currently, although the number densities 
will be growing. For instance, {\it CosmicFlows-4} is expected to contain of the order of 
$3\times10^4$ sources but still mostly within $R<150\hmpc$ as currently 
CosmicFlows-3 does\footnote{Tully, private communication.}. 
It is only the advent of all-sky HI radio surveys that can extend 
the reach of PV surveys to $\sim 2$ times larger distances, and the object number closer to $10^5$.

Careful modelling of observer location,
and survey selection strategy are necessary for obtaining 
reliable and unbiased velocity correlation estimates. Much more effort is required to 
extract cosmological information richly stored in galaxy velocity data.
Towards this goal, constrained realization techniques 
\citep[][]{HoffmanRibak1991,vdWeygaert1996,Klypin2003,Courtois2012,Hess2013,Sorce2016},
aiming at incorporating prominent structures in the real Universe can be very rewarding.

\section*{Acknowledgements}
The Authors have benefited from many inspiring discussions with
H\'{e}l\`{e}ne Curtois, Noam Libeskind and Brent Tully.
The hospitality of the Technion -- Israel Institute of Technology
enjoyed by WAH,MF and MB at the early stages of this project is warmly acknowledged. 
WAH acknowledges support from European Research Council 
(grant number 646702 ``CosTesGrav''). 
WAH and MB were also supported by the Polish National Science Center under contract \#UMO-2012/07/D/ST9/02785. 
This research was supported by the I-CORE Program of the Planning and Budgeting Committee, 
THE ISRAEL SCIENCE FOUNDATION (grants No. 1829/12 and No. 203/09) and the Asher Space Research Institute.
MF acknowledges support through the Grant Spin(e) ANR-13-BS05-0005 of the French National Research Agency.
\label{sec:appendix}

\bibliographystyle{mnras}
\bibliography{vel_stats}

\begin{thebibliography}{}
\makeatletter
\relax
\def\mn@urlcharsother{\let\do\@makeother \do\$\do\&\do\#\do\^\do\_\do\%\do\~}
\def\mn@doi{\begingroup\mn@urlcharsother \@ifnextchar [ {\mn@doi@}
  {\mn@doi@[]}}
\def\mn@doi@[#1]#2{\def\@tempa{#1}\ifx\@tempa\@empty \href
  {http://dx.doi.org/#2} {doi:#2}\else \href {http://dx.doi.org/#2} {#1}\fi
  \endgroup}
\def\mn@eprint#1#2{\mn@eprint@#1:#2::\@nil}
\def\mn@eprint@arXiv#1{\href {http://arxiv.org/abs/#1} {{\tt arXiv:#1}}}
\def\mn@eprint@dblp#1{\href {http://dblp.uni-trier.de/rec/bibtex/#1.xml}
  {dblp:#1}}
\def\mn@eprint@#1:#2:#3:#4\@nil{\def\@tempa {#1}\def\@tempb {#2}\def\@tempc
  {#3}\ifx \@tempc \@empty \let \@tempc \@tempb \let \@tempb \@tempa \fi \ifx
  \@tempb \@empty \def\@tempb {arXiv}\fi \@ifundefined
  {mn@eprint@\@tempb}{\@tempb:\@tempc}{\expandafter \expandafter \csname
  mn@eprint@\@tempb\endcsname \expandafter{\@tempc}}}

\bibitem[\protect\citeauthoryear{{Alpaslan} et~al.,}{{Alpaslan}
  et~al.}{2014}]{Alpaslan2014}
{Alpaslan} M.,  et~al., 2014, \mn@doi [\mnras] {10.1093/mnras/stt2136}, \href
  {http://adsabs.harvard.edu/abs/2014MNRAS.438..177A} {438, 177}

\bibitem[\protect\citeauthoryear{{Behroozi}, {Wechsler}  \& {Wu}}{{Behroozi}
  et~al.}{2013}]{Behroozi2013}
{Behroozi} P.~S.,  {Wechsler} R.~H.,   {Wu} H.-Y.,  2013, \mn@doi [\apj]
  {10.1088/0004-637X/762/2/109}, \href
  {http://adsabs.harvard.edu/abs/2013ApJ...762..109B} {762, 109}

\bibitem[\protect\citeauthoryear{{Berti} et~al.,}{{Berti}
  et~al.}{2015}]{Berti2015}
{Berti} E.,  et~al., 2015, \mn@doi [Classical and Quantum Gravity]
  {10.1088/0264-9381/32/24/243001}, \href
  {http://adsabs.harvard.edu/abs/2015CQGra..32x3001B} {32, 243001}

\bibitem[\protect\citeauthoryear{{Bilicki} \& {Chodorowski}}{{Bilicki} \&
  {Chodorowski}}{2010}]{Bilicki2010}
{Bilicki} M.,  {Chodorowski} M.~J.,  2010, \mn@doi [\mnras]
  {10.1111/j.1365-2966.2010.16771.x}, \href
  {http://esoads.eso.org/abs/2010MNRAS.406.1358B} {406, 1358}

\bibitem[\protect\citeauthoryear{{Borgani}, {da Costa}, {Zehavi}, {Giovanelli},
  {Haynes}, {Freudling}, {Wegner}  \& {Salzer}}{{Borgani}
  et~al.}{2000}]{Borgani2000}
{Borgani} S.,  {da Costa} L.~N.,  {Zehavi} I.,  {Giovanelli} R.,  {Haynes}
  M.~P.,  {Freudling} W.,  {Wegner} G.,   {Salzer} J.~J.,  2000, \mn@doi [\aj]
  {10.1086/301154}, \href {http://adsabs.harvard.edu/abs/2000AJ....119..102B}
  {119, 102}

\bibitem[\protect\citeauthoryear{{Bull}}{{Bull}}{2016}]{Bull2016}
{Bull} P.,  2016, \mn@doi [\apj] {10.3847/0004-637X/817/1/26}, \href
  {http://adsabs.harvard.edu/abs/2016ApJ...817...26B} {817, 26}

\bibitem[\protect\citeauthoryear{{Busha}, {Wechsler}, {Behroozi}, {Gerke},
  {Klypin}  \& {Primack}}{{Busha} et~al.}{2011}]{Busha2011}
{Busha} M.~T.,  {Wechsler} R.~H.,  {Behroozi} P.~S.,  {Gerke} B.~F.,  {Klypin}
  A.~A.,   {Primack} J.~R.,  2011, \mn@doi [\apj]
  {10.1088/0004-637X/743/2/117}, \href
  {http://esoads.eso.org/abs/2011ApJ...743..117B} {743, 117}

\bibitem[\protect\citeauthoryear{{Cautun} \& {van de Weygaert}}{{Cautun} \&
  {van de Weygaert}}{2011}]{cv2011}
{Cautun} M.~C.,  {van de Weygaert} R.,  2011, preprint, \href
  {http://adsabs.harvard.edu/abs/2011arXiv1105.0370C} {} (\mn@eprint {arXiv}
  {1105.0370})

\bibitem[\protect\citeauthoryear{{Cautun}, {Frenk}, {van de Weygaert},
  {Hellwing}  \& {Jones}}{{Cautun} et~al.}{2014}]{Cautun2014}
{Cautun} M.,  {Frenk} C.~S.,  {van de Weygaert} R.,  {Hellwing} W.~A.,
  {Jones} B.~J.~T.,  2014, \mn@doi [\mnras] {10.1093/mnras/stu1849}, \href
  {http://adsabs.harvard.edu/abs/2014MNRAS.445.2049C} {445, 2049}

\bibitem[\protect\citeauthoryear{{Chodorowski} \& {Ciecielag}}{{Chodorowski} \&
  {Ciecielag}}{2002}]{Chodorowski2002}
{Chodorowski} M.~J.,  {Ciecielag} P.,  2002, \mn@doi [\mnras]
  {10.1046/j.1365-8711.2002.05161.x}, \href
  {http://ukads.nottingham.ac.uk/abs/2002MNRAS.331..133C} {331, 133}

\bibitem[\protect\citeauthoryear{{Colless} et~al.,}{{Colless}
  et~al.}{2001}]{2dfgrs}
{Colless} M.,  et~al., 2001, \mn@doi [\mnras]
  {10.1046/j.1365-8711.2001.04902.x}, \href
  {http://adsabs.harvard.edu/abs/2001MNRAS.328.1039C} {328, 1039}

\bibitem[\protect\citeauthoryear{{Cooray} \& {Caldwell}}{{Cooray} \&
  {Caldwell}}{2006}]{Cooray2006}
{Cooray} A.,  {Caldwell} R.~R.,  2006, \mn@doi [\prd]
  {10.1103/PhysRevD.73.103002}, \href
  {http://esoads.eso.org/abs/2006PhRvD..73j3002C} {73, 103002}

\bibitem[\protect\citeauthoryear{{Courtois} \& {Tully}}{{Courtois} \&
  {Tully}}{2012}]{Courtois2012}
{Courtois} H.~M.,  {Tully} R.~B.,  2012, \mn@doi [Astronomische Nachrichten]
  {10.1002/asna.201211682}, \href
  {http://adsabs.harvard.edu/abs/2012AN....333..436C} {333, 436}

\bibitem[\protect\citeauthoryear{{Courtois}, {Tully}, {Makarov}, {Mitronova},
  {Koribalski}, {Karachentsev}  \& {Fisher}}{{Courtois}
  et~al.}{2011}]{Courtois2011}
{Courtois} H.~M.,  {Tully} R.~B.,  {Makarov} D.~I.,  {Mitronova} S.,
  {Koribalski} B.,  {Karachentsev} I.~D.,   {Fisher} J.~R.,  2011, \mn@doi
  [\mnras] {10.1111/j.1365-2966.2011.18515.x}, \href
  {http://adsabs.harvard.edu/abs/2011MNRAS.414.2005C} {414, 2005}

\bibitem[\protect\citeauthoryear{{Courtois}, {Pomar{\`e}de}, {Tully}, {Hoffman}
   \& {Courtois}}{{Courtois} et~al.}{2013}]{Courtois2013}
{Courtois} H.~M.,  {Pomar{\`e}de} D.,  {Tully} R.~B.,  {Hoffman} Y.,
  {Courtois} D.,  2013, \mn@doi [\aj] {10.1088/0004-6256/146/3/69}, \href
  {http://esoads.eso.org/abs/2013AJ....146...69C} {146, 69}

\bibitem[\protect\citeauthoryear{{Davis} \& {Peebles}}{{Davis} \&
  {Peebles}}{1977}]{DavisPeebles_BBGKY}
{Davis} M.,  {Peebles} P.~J.~E.,  1977, \mn@doi [\apjs] {10.1086/190456}, \href
  {http://adsabs.harvard.edu/abs/1977ApJS...34..425D} {34, 425}

\bibitem[\protect\citeauthoryear{{Davis}, {Nusser}, {Masters}, {Springob},
  {Huchra}  \& {Lemson}}{{Davis} et~al.}{2011}]{Davis2011}
{Davis} M.,  {Nusser} A.,  {Masters} K.~L.,  {Springob} C.,  {Huchra} J.~P.,
  {Lemson} G.,  2011, \mn@doi [\mnras] {10.1111/j.1365-2966.2011.18362.x},
  \href {http://adsabs.harvard.edu/abs/2011MNRAS.413.2906D} {413, 2906}

\bibitem[\protect\citeauthoryear{{Elyiv}, {Karachentsev}, {Karachentseva},
  {Melnyk}  \& {Makarov}}{{Elyiv} et~al.}{2013}]{Elyiv2013}
{Elyiv} A.~A.,  {Karachentsev} I.~D.,  {Karachentseva} V.~E.,  {Melnyk} O.~V.,
   {Makarov} D.~I.,  2013, \mn@doi [Astrophysical Bulletin]
  {10.1134/S199034131301001X}, \href
  {http://esoads.eso.org/abs/2013AstBu..68....1E} {68, 1}

\bibitem[\protect\citeauthoryear{{Fagernes Ivarsen}, {Bull}, {Llinares}  \&
  {Mota}}{{Fagernes Ivarsen} et~al.}{2016}]{Ivarsen2016}
{Fagernes Ivarsen} M.,  {Bull} P.,  {Llinares} C.,   {Mota} D.~F.,  2016,
  preprint, \href {http://adsabs.harvard.edu/abs/2016arXiv160303072F} {}
  (\mn@eprint {arXiv} {1603.03072})

\bibitem[\protect\citeauthoryear{{Feix}, {Nusser}  \& {Branchini}}{{Feix}
  et~al.}{2015}]{Feix2015}
{Feix} M.,  {Nusser} A.,   {Branchini} E.,  2015, \mn@doi [Physical Review
  Letters] {10.1103/PhysRevLett.115.011301}, \href
  {http://adsabs.harvard.edu/abs/2015PhRvL.115a1301F} {115, 011301}

\bibitem[\protect\citeauthoryear{{Feldman} et~al.,}{{Feldman}
  et~al.}{2003a}]{streaming_vel_Omega}
{Feldman} H.,  et~al., 2003a, \mn@doi [\apjl] {10.1086/379221}, \href
  {http://adsabs.harvard.edu/abs/2003ApJ...596L.131F} {596, L131}

\bibitem[\protect\citeauthoryear{{Feldman} et~al.,}{{Feldman}
  et~al.}{2003b}]{Feldman2003}
{Feldman} H.,  et~al., 2003b, \mn@doi [\apjl] {10.1086/379221}, \href
  {http://adsabs.harvard.edu/abs/2003ApJ...596L.131F} {596, L131}

\bibitem[\protect\citeauthoryear{{Ferreira}, {Juszkiewicz}, {Feldman}, {Davis}
  \& {Jaffe}}{{Ferreira} et~al.}{1999}]{Ferreira1999}
{Ferreira} P.~G.,  {Juszkiewicz} R.,  {Feldman} H.~A.,  {Davis} M.,   {Jaffe}
  A.~H.,  1999, \mn@doi [\apjl] {10.1086/311959}, \href
  {http://ukads.nottingham.ac.uk/abs/1999ApJ...515L...1F} {515, L1}

\bibitem[\protect\citeauthoryear{{Geller} \& {Huchra}}{{Geller} \&
  {Huchra}}{1989}]{Geller1989}
{Geller} M.~J.,  {Huchra} J.~P.,  1989, \mn@doi [Science]
  {10.1126/science.246.4932.897}, \href
  {http://esoads.eso.org/abs/1989Sci...246..897G} {246, 897}

\bibitem[\protect\citeauthoryear{{G\'orski}}{{G\'orski}}{1988}]{Gorski1988}
{G\'orski} K.,  1988, \mn@doi [\apjl] {10.1086/185255}, \href
  {http://ukads.nottingham.ac.uk/abs/1988ApJ...332L...7G} {332, L7}

\bibitem[\protect\citeauthoryear{{G\'orski}, {Davis}, {Strauss}, {White}  \&
  {Yahil}}{{G\'orski} et~al.}{1989}]{Gorski1989}
{G\'orski} K.~M.,  {Davis} M.,  {Strauss} M.~A.,  {White} S.~D.~M.,   {Yahil}
  A.,  1989, \mn@doi [\apj] {10.1086/167771}, \href
  {http://ukads.nottingham.ac.uk/abs/1989ApJ...344....1G} {344, 1}

\bibitem[\protect\citeauthoryear{{Groth}, {Juszkiewicz}  \& {Ostriker}}{{Groth}
  et~al.}{1989}]{Groth1989}
{Groth} E.~J.,  {Juszkiewicz} R.,   {Ostriker} J.~P.,  1989, \mn@doi [\apj]
  {10.1086/168038}, \href {http://esoads.eso.org/abs/1989ApJ...346..558G} {346,
  558}

\bibitem[\protect\citeauthoryear{{Gudehus}}{{Gudehus}}{1995}]{Gudehus1995}
{Gudehus} D.,  1995, \aap, \href
  {http://esoads.eso.org/abs/1995A%26A...302...21G} {302, 21}

\bibitem[\protect\citeauthoryear{{Guo}, {Cooper}, {Frenk}, {Helly}  \&
  {Hellwing}}{{Guo} et~al.}{2015}]{GuoCooper2015}
{Guo} Q.,  {Cooper} A.~P.,  {Frenk} C.,  {Helly} J.,   {Hellwing} W.~A.,  2015,
  \mn@doi [\mnras] {10.1093/mnras/stv1938}, \href
  {http://adsabs.harvard.edu/abs/2015MNRAS.454..550G} {454, 550}

\bibitem[\protect\citeauthoryear{{Guzzo} et~al.,}{{Guzzo}
  et~al.}{2014}]{Guzzo2014}
{Guzzo} L.,  et~al., 2014, \mn@doi [\aap] {10.1051/0004-6361/201321489}, \href
  {http://adsabs.harvard.edu/abs/2014A%26A...566A.108G} {566, A108}

\bibitem[\protect\citeauthoryear{{Hellwing}}{{Hellwing}}{2014}]{Hellwing2014}
{Hellwing} W.~A.,  2014, preprint, \href
  {http://adsabs.harvard.edu/abs/2014arXiv1412.8738H} {} (\mn@eprint {arXiv}
  {1412.8738})

\bibitem[\protect\citeauthoryear{{Hellwing}, {Barreira}, {Frenk}, {Li}  \&
  {Cole}}{{Hellwing} et~al.}{2014}]{Hellwing2014PhRvL}
{Hellwing} W.~A.,  {Barreira} A.,  {Frenk} C.~S.,  {Li} B.,   {Cole} S.,  2014,
  \mn@doi [Physical Review Letters] {10.1103/PhysRevLett.112.221102}, \href
  {http://adsabs.harvard.edu/abs/2014PhRvL.112v1102H} {112, 221102}

\bibitem[\protect\citeauthoryear{{Hellwing}, {Schaller}, {Frenk}, {Theuns},
  {Schaye}, {Bower}  \& {Crain}}{{Hellwing} et~al.}{2016}]{Hellwing2016}
{Hellwing} W.~A.,  {Schaller} M.,  {Frenk} C.~S.,  {Theuns} T.,  {Schaye} J.,
  {Bower} R.~G.,   {Crain} R.~A.,  2016, \mn@doi [\mnras]
  {10.1093/mnrasl/slw081}, \href
  {http://adsabs.harvard.edu/abs/2016MNRAS.461L..11H} {461, L11}

\bibitem[\protect\citeauthoryear{{He{\ss}}, {Kitaura}  \&
  {Gottl{\"o}ber}}{{He{\ss}} et~al.}{2013}]{Hess2013}
{He{\ss}} S.,  {Kitaura} F.-S.,   {Gottl{\"o}ber} S.,  2013, \mn@doi [\mnras]
  {10.1093/mnras/stt1428}, \href
  {http://adsabs.harvard.edu/abs/2013MNRAS.435.2065H} {435, 2065}

\bibitem[\protect\citeauthoryear{{Hinshaw} et~al.,}{{Hinshaw}
  et~al.}{2013}]{WMAP9}
{Hinshaw} G.,  et~al., 2013, \mn@doi [\apjs] {10.1088/0067-0049/208/2/19},
  \href {http://adsabs.harvard.edu/abs/2013ApJS..208...19H} {208, 19}

\bibitem[\protect\citeauthoryear{{Hoffman} \& {Ribak}}{{Hoffman} \&
  {Ribak}}{1991}]{HoffmanRibak1991}
{Hoffman} Y.,  {Ribak} E.,  1991, \mn@doi [\apjl] {10.1086/186160}, \href
  {http://adsabs.harvard.edu/abs/1991ApJ...380L...5H} {380, L5}

\bibitem[\protect\citeauthoryear{{Huchra}, {Davis}, {Latham}  \&
  {Tonry}}{{Huchra} et~al.}{1983}]{Huchra1983}
{Huchra} J.,  {Davis} M.,  {Latham} D.,   {Tonry} J.,  1983, \mn@doi [\apjs]
  {10.1086/190860}, \href {http://esoads.eso.org/abs/1983ApJS...52...89H} {52,
  89}

\bibitem[\protect\citeauthoryear{{Hudson}}{{Hudson}}{1993}]{Hudson1993}
{Hudson} M.~J.,  1993, \mn@doi [\mnras] {10.1093/mnras/265.1.43}, \href
  {http://esoads.eso.org/abs/1993MNRAS.265...43H} {265, 43}

\bibitem[\protect\citeauthoryear{{Hudson} \& {Turnbull}}{{Hudson} \&
  {Turnbull}}{2012}]{Hudson2012}
{Hudson} M.~J.,  {Turnbull} S.~J.,  2012, \mn@doi [\apjl]
  {10.1088/2041-8205/751/2/L30}, \href
  {http://adsabs.harvard.edu/abs/2012ApJ...751L..30H} {751, L30}

\bibitem[\protect\citeauthoryear{{Juszkiewicz}, {Springel}  \&
  {Durrer}}{{Juszkiewicz} et~al.}{1999}]{Juszkiewicz1999}
{Juszkiewicz} R.,  {Springel} V.,   {Durrer} R.,  1999, \mn@doi [\apjl]
  {10.1086/312055}, \href {http://adsabs.harvard.edu/abs/1999ApJ...518L..25J}
  {518, L25}

\bibitem[\protect\citeauthoryear{{Juszkiewicz}, {Ferreira}, {Feldman}, {Jaffe}
  \& {Davis}}{{Juszkiewicz} et~al.}{2000}]{Juszkiewicz2000}
{Juszkiewicz} R.,  {Ferreira} P.~G.,  {Feldman} H.~A.,  {Jaffe} A.~H.,
  {Davis} M.,  2000, \mn@doi [Science] {10.1126/science.287.5450.109}, \href
  {http://adsabs.harvard.edu/abs/2000Sci...287..109J} {287, 109}

\bibitem[\protect\citeauthoryear{{Karachentsev} et~al.,}{{Karachentsev}
  et~al.}{2002}]{Karachentsev2002}
{Karachentsev} I.~D.,  et~al., 2002, \mn@doi [\aap]
  {10.1051/0004-6361:20020649}, \href
  {http://esoads.eso.org/abs/2002A%26A...389..812K} {389, 812}

\bibitem[\protect\citeauthoryear{{Karachentsev} et~al.,}{{Karachentsev}
  et~al.}{2003}]{Karachentsev2003}
{Karachentsev} I.~D.,  et~al., 2003, \mn@doi [\aap]
  {10.1051/0004-6361:20021566}, \href
  {http://esoads.eso.org/abs/2003A%26A...398..479K} {398, 479}

\bibitem[\protect\citeauthoryear{{Karachentsev}, {Karachentseva}, {Melnyk},
  {Elyiv}  \& {Makarov}}{{Karachentsev} et~al.}{2012}]{Karachentsev2012}
{Karachentsev} I.~D.,  {Karachentseva} V.~E.,  {Melnyk} O.~V.,  {Elyiv} A.~A.,
   {Makarov} D.~I.,  2012, \mn@doi [Astrophysical Bulletin]
  {10.1134/S1990341312040013}, \href
  {http://esoads.eso.org/abs/2012AstBu..67..353K} {67, 353}

\bibitem[\protect\citeauthoryear{{Karachentsev}, {Tully}, {Wu}, {Shaya}  \&
  {Dolphin}}{{Karachentsev} et~al.}{2014}]{Karachentsev2014}
{Karachentsev} I.~D.,  {Tully} R.~B.,  {Wu} P.-F.,  {Shaya} E.~J.,   {Dolphin}
  A.~E.,  2014, \mn@doi [\apj] {10.1088/0004-637X/782/1/4}, \href
  {http://esoads.eso.org/abs/2014ApJ...782....4K} {782, 4}

\bibitem[\protect\citeauthoryear{{Klypin}, {Hoffman}, {Kravtsov}  \&
  {Gottl{\"o}ber}}{{Klypin} et~al.}{2003}]{Klypin2003}
{Klypin} A.,  {Hoffman} Y.,  {Kravtsov} A.~V.,   {Gottl{\"o}ber} S.,  2003,
  \mn@doi [\apj] {10.1086/377574}, \href
  {http://adsabs.harvard.edu/abs/2003ApJ...596...19K} {596, 19}

\bibitem[\protect\citeauthoryear{{Kogut} et~al.,}{{Kogut}
  et~al.}{1993}]{Kogut1993}
{Kogut} A.,  et~al., 1993, \mn@doi [\apj] {10.1086/173453}, \href
  {http://esoads.eso.org/abs/1993ApJ...419....1K} {419, 1}

\bibitem[\protect\citeauthoryear{{Komatsu} \& {et al.}}{{Komatsu} \& {et
  al.}}{2010}]{WMAP7}
{Komatsu} E.,  {et al.} 2010, preprint, \href
  {http://adsabs.harvard.edu/abs/2010arXiv1001.4538K} {} (\mn@eprint {arXiv}
  {1001.4538})

\bibitem[\protect\citeauthoryear{{Li}, {Hellwing}, {Koyama}, {Zhao}, {Jennings}
   \& {Baugh}}{{Li} et~al.}{2013}]{Li2013}
{Li} B.,  {Hellwing} W.~A.,  {Koyama} K.,  {Zhao} G.-B.,  {Jennings} E.,
  {Baugh} C.~M.,  2013, \mn@doi [\mnras] {10.1093/mnras/sts072}, \href
  {http://adsabs.harvard.edu/abs/2013MNRAS.428..743L} {428, 743}

\bibitem[\protect\citeauthoryear{{Lu}, {Salpeter}  \& {Hoffman}}{{Lu}
  et~al.}{1994}]{Lu1994}
{Lu} N.~Y.,  {Salpeter} E.~E.,   {Hoffman} G.~L.,  1994, \mn@doi [\apj]
  {10.1086/174083}, \href {http://esoads.eso.org/abs/1994ApJ...426..473L} {426,
  473}

\bibitem[\protect\citeauthoryear{{Lynden-Bell}, {Burstein}, {Davies},
  {Dressler}  \& {Faber}}{{Lynden-Bell} et~al.}{1988a}]{Lynden-Bell1988a}
{Lynden-Bell} D.,  {Burstein} D.,  {Davies} R.~L.,  {Dressler} A.,   {Faber}
  S.~M.,  1988a, in {van den Bergh} S.,  {Pritchet} C.~J.,  eds,  Astronomical
  Society of the Pacific Conference Series Vol. 4, The Extragalactic Distance
  Scale. pp 307--316

\bibitem[\protect\citeauthoryear{{Lynden-Bell}, {Faber}, {Burstein}, {Davies},
  {Dressler}, {Terlevich}  \& {Wegner}}{{Lynden-Bell}
  et~al.}{1988b}]{Lynden-Bell1988b}
{Lynden-Bell} D.,  {Faber} S.~M.,  {Burstein} D.,  {Davies} R.~L.,  {Dressler}
  A.,  {Terlevich} R.~J.,   {Wegner} G.,  1988b, \mn@doi [\apj]
  {10.1086/166066}, \href {http://esoads.eso.org/abs/1988ApJ...326...19L} {326,
  19}

\bibitem[\protect\citeauthoryear{{Ma}, {Li}  \& {He}}{{Ma}
  et~al.}{2015}]{Ma2015}
{Ma} Y.-Z.,  {Li} M.,   {He} P.,  2015, \mn@doi [\aap]
  {10.1051/0004-6361/201526051}, \href
  {http://adsabs.harvard.edu/abs/2015A%26A...583A..52M} {583, A52}

\bibitem[\protect\citeauthoryear{{Marra}, {Amendola}, {Sawicki}  \&
  {Valkenburg}}{{Marra} et~al.}{2013}]{Marra2013}
{Marra} V.,  {Amendola} L.,  {Sawicki} I.,   {Valkenburg} W.,  2013, \mn@doi
  [Physical Review Letters] {10.1103/PhysRevLett.110.241305}, \href
  {http://esoads.eso.org/abs/2013PhRvL.110x1305M} {110, 241305}

\bibitem[\protect\citeauthoryear{{Mei} et~al.,}{{Mei} et~al.}{2007}]{Mei2007}
{Mei} S.,  et~al., 2007, \mn@doi [\apj] {10.1086/509598}, \href
  {http://adsabs.harvard.edu/abs/2007ApJ...655..144M} {655, 144}

\bibitem[\protect\citeauthoryear{{Milne}}{{Milne}}{1935}]{Milne1935}
{Milne} E.~A.,  1935, {Relativity, gravitation and world-structure}

\bibitem[\protect\citeauthoryear{{Nadathur}}{{Nadathur}}{2013}]{Nadathur2013}
{Nadathur} S.,  2013, \mn@doi [\mnras] {10.1093/mnras/stt1028}, \href
  {http://adsabs.harvard.edu/abs/2013MNRAS.434..398N} {434, 398}

\bibitem[\protect\citeauthoryear{{Nusser} \& {Colberg}}{{Nusser} \&
  {Colberg}}{1998}]{Nusser1998}
{Nusser} A.,  {Colberg} J.~M.,  1998, \mn@doi [\mnras]
  {10.1046/j.1365-8711.1998.01218.x}, \href
  {http://ukads.nottingham.ac.uk/abs/1998MNRAS.294..457N} {294, 457}

\bibitem[\protect\citeauthoryear{{Nusser} \& {Lahav}}{{Nusser} \&
  {Lahav}}{2000}]{NusserLahav2000}
{Nusser} A.,  {Lahav} O.,  2000, \mn@doi [\mnras]
  {10.1046/j.1365-8711.2000.03441.x}, \href
  {http://adsabs.harvard.edu/abs/2000MNRAS.313L..39N} {313, L39}

\bibitem[\protect\citeauthoryear{{Nusser}, {Branchini}  \& {Davis}}{{Nusser}
  et~al.}{2011}]{Nusser2011}
{Nusser} A.,  {Branchini} E.,   {Davis} M.,  2011, \mn@doi [\apj]
  {10.1088/0004-637X/735/2/77}, \href
  {http://adsabs.harvard.edu/abs/2011ApJ...735...77N} {735, 77}

\bibitem[\protect\citeauthoryear{{Nusser}, {Branchini}  \& {Davis}}{{Nusser}
  et~al.}{2012}]{Nusser2012}
{Nusser} A.,  {Branchini} E.,   {Davis} M.,  2012, \mn@doi [\apj]
  {10.1088/0004-637X/755/1/58}, \href
  {http://adsabs.harvard.edu/abs/2012ApJ...755...58N} {755, 58}

\bibitem[\protect\citeauthoryear{{Peebles}}{{Peebles}}{1980}]{1980Peebles}
{Peebles} P.~J.~E.,  1980, {The large-scale structure of the universe}.
Research supported by the National Science Foundation.~Princeton, N.J.,
  Princeton University Press, 1980.~435 p.

\bibitem[\protect\citeauthoryear{{Phelps}, {Nusser}  \& {Desjacques}}{{Phelps}
  et~al.}{2013}]{Phelps2013}
{Phelps} S.,  {Nusser} A.,   {Desjacques} V.,  2013, \mn@doi [\apj]
  {10.1088/0004-637X/775/2/102}, \href
  {http://adsabs.harvard.edu/abs/2013ApJ...775..102P} {775, 102}

\bibitem[\protect\citeauthoryear{{Planck Collaboration} et~al.,}{{Planck
  Collaboration} et~al.}{2015}]{Planck2015}
{Planck Collaboration} et~al., 2015, preprint, \href
  {http://adsabs.harvard.edu/abs/2015arXiv150201589P} {} (\mn@eprint {arXiv}
  {1502.01589})

\bibitem[\protect\citeauthoryear{{S{\'a}nchez} et~al.,}{{S{\'a}nchez}
  et~al.}{2012}]{CMASS_corr2012}
{S{\'a}nchez} A.~G.,  et~al., 2012, \mn@doi [\mnras]
  {10.1111/j.1365-2966.2012.21502.x}, \href
  {http://adsabs.harvard.edu/abs/2012MNRAS.425..415S} {425, 415}

\bibitem[\protect\citeauthoryear{{Sandage}}{{Sandage}}{1978}]{Sandage1978}
{Sandage} A.,  1978, \mn@doi [\aj] {10.1086/112271}, \href
  {http://esoads.eso.org/abs/1978AJ.....83..904S} {83, 904}

\bibitem[\protect\citeauthoryear{{Sandage}, {Tammann}  \& {Hardy}}{{Sandage}
  et~al.}{1972}]{Sandage1972}
{Sandage} A.,  {Tammann} G.~A.,   {Hardy} E.,  1972, \mn@doi [\apj]
  {10.1086/151344}, \href {http://esoads.eso.org/abs/1972ApJ...172..253S} {172,
  253}

\bibitem[\protect\citeauthoryear{{Sarkar}, {Feldman}  \& {Watkins}}{{Sarkar}
  et~al.}{2007}]{Sarkar2007}
{Sarkar} D.,  {Feldman} H.~A.,   {Watkins} R.,  2007, \mn@doi [\mnras]
  {10.1111/j.1365-2966.2006.11334.x}, \href
  {http://adsabs.harvard.edu/abs/2007MNRAS.375..691S} {375, 691}

\bibitem[\protect\citeauthoryear{{Schaap} \& {van de Weygaert}}{{Schaap} \&
  {van de Weygaert}}{2000}]{sv2000}
{Schaap} W.~E.,  {van de Weygaert} R.,  2000, \aap, \href
  {http://adsabs.harvard.edu/abs/2000A%26A...363L..29S} {363, L29}

\bibitem[\protect\citeauthoryear{{Schaye} et~al.,}{{Schaye}
  et~al.}{2015}]{Schaye2015}
{Schaye} J.,  et~al., 2015, \mn@doi [\mnras] {10.1093/mnras/stu2058}, \href
  {http://adsabs.harvard.edu/abs/2015MNRAS.446..521S} {446, 521}

\bibitem[\protect\citeauthoryear{{Schlegel}, {Davis}, {Summers}  \&
  {Holtzman}}{{Schlegel} et~al.}{1994}]{Schlegel1994}
{Schlegel} D.,  {Davis} M.,  {Summers} F.,   {Holtzman} J.~A.,  1994, \mn@doi
  [\apj] {10.1086/174164}, \href
  {http://adsabs.harvard.edu/abs/1994ApJ...427..527S} {427, 527}

\bibitem[\protect\citeauthoryear{{Scrimgeour} et~al.,}{{Scrimgeour}
  et~al.}{2012}]{Scrimgeour2012}
{Scrimgeour} M.~I.,  et~al., 2012, \mn@doi [\mnras]
  {10.1111/j.1365-2966.2012.21402.x}, \href
  {http://adsabs.harvard.edu/abs/2012MNRAS.425..116S} {425, 116}

\bibitem[\protect\citeauthoryear{{Soifer} et~al.,}{{Soifer}
  et~al.}{1984}]{Soifer1984}
{Soifer} B.~T.,  et~al., 1984, \mn@doi [\apjl] {10.1086/184226}, \href
  {http://esoads.eso.org/abs/1984ApJ...278L..71S} {278, L71}

\bibitem[\protect\citeauthoryear{{Sorce} et~al.,}{{Sorce}
  et~al.}{2016}]{Sorce2016}
{Sorce} J.~G.,  et~al., 2016, \mn@doi [\mnras] {10.1093/mnras/stv2407}, \href
  {http://adsabs.harvard.edu/abs/2016MNRAS.455.2078S} {455, 2078}

\bibitem[\protect\citeauthoryear{{Springel}}{{Springel}}{2005}]{Gadget2}
{Springel} V.,  2005, \mn@doi [\mnras] {10.1111/j.1365-2966.2005.09655.x},
  \href {http://ads.nao.ac.jp/abs/2005MNRAS.364.1105S} {364, 1105}

\bibitem[\protect\citeauthoryear{{Springob}, {Masters}, {Haynes}, {Giovanelli}
  \& {Marinoni}}{{Springob} et~al.}{2007}]{Springob2007}
{Springob} C.~M.,  {Masters} K.~L.,  {Haynes} M.~P.,  {Giovanelli} R.,
  {Marinoni} C.,  2007, \mn@doi [\apjs] {10.1086/519527}, \href
  {http://esoads.eso.org/abs/2007ApJS..172..599S} {172, 599}

\bibitem[\protect\citeauthoryear{{Springob} et~al.,}{{Springob}
  et~al.}{2014}]{Springob2014}
{Springob} C.~M.,  et~al., 2014, \mn@doi [\mnras] {10.1093/mnras/stu1743},
  \href {http://esoads.eso.org/abs/2014MNRAS.445.2677S} {445, 2677}

\bibitem[\protect\citeauthoryear{{Strauss} \& {Willick}}{{Strauss} \&
  {Willick}}{1995}]{StraussWillick}
{Strauss} M.~A.,  {Willick} J.~A.,  1995, \mn@doi [\physrep]
  {10.1016/0370-1573(95)00013-7}, \href
  {http://ukads.nottingham.ac.uk/abs/1995PhR...261..271S} {261, 271}

\bibitem[\protect\citeauthoryear{{Strauss}, {Ostriker}  \& {Cen}}{{Strauss}
  et~al.}{1998}]{Strauss1998}
{Strauss} M.~A.,  {Ostriker} J.~P.,   {Cen} R.,  1998, \mn@doi [\apj]
  {10.1086/305211}, \href
  {http://ukads.nottingham.ac.uk/abs/1998ApJ...494...20S} {494, 20}

\bibitem[\protect\citeauthoryear{{Tammann} \& {Sandage}}{{Tammann} \&
  {Sandage}}{1985}]{Tammann1985}
{Tammann} G.~A.,  {Sandage} A.,  1985, \mn@doi [\apj] {10.1086/163277}, \href
  {http://esoads.eso.org/abs/1985ApJ...294...81T} {294, 81}

\bibitem[\protect\citeauthoryear{{Tegmark} et~al.,}{{Tegmark}
  et~al.}{2004}]{sdss}
{Tegmark} M.,  et~al., 2004, \mn@doi [\prd] {10.1103/PhysRevD.69.103501}, \href
  {http://adsabs.harvard.edu/abs/2004PhRvD..69j3501T} {69, 103501}

\bibitem[\protect\citeauthoryear{{The Planck Collaboration} et~al.,}{{The
  Planck Collaboration} et~al.}{2013}]{Planck2013}
{The Planck Collaboration} et~al., 2013, preprint, \href
  {http://adsabs.harvard.edu/abs/2013arXiv1303.5076T} {} (\mn@eprint {arXiv}
  {1303.5076})

\bibitem[\protect\citeauthoryear{{Tormen}, {Moscardini}, {Lucchin}  \&
  {Matarrese}}{{Tormen} et~al.}{1993}]{Tormen1993}
{Tormen} G.,  {Moscardini} L.,  {Lucchin} F.,   {Matarrese} S.,  1993, \mn@doi
  [\apj] {10.1086/172804}, \href
  {http://adsabs.harvard.edu/abs/1993ApJ...411...16T} {411, 16}

\bibitem[\protect\citeauthoryear{{Tully} \& {Fisher}}{{Tully} \&
  {Fisher}}{1987}]{Tully1987}
{Tully} R.~B.,  {Fisher} J.~R.,  1987, {Atlas of Nearby Galaxies}

\bibitem[\protect\citeauthoryear{{Tully} \& {Fisher}}{{Tully} \&
  {Fisher}}{1988}]{Tully1988}
{Tully} R.~B.,  {Fisher} J.~R.,  1988, {Catalog of Nearby Galaxies}

\bibitem[\protect\citeauthoryear{{Tully} \& {Shaya}}{{Tully} \&
  {Shaya}}{1984}]{TullyShaya1984}
{Tully} R.~B.,  {Shaya} E.~J.,  1984, \mn@doi [\apj] {10.1086/162073}, \href
  {http://esoads.eso.org/abs/1984ApJ...281...31T} {281, 31}

\bibitem[\protect\citeauthoryear{{Tully}, {Shaya}, {Karachentsev}, {Courtois},
  {Kocevski}, {Rizzi}  \& {Peel}}{{Tully} et~al.}{2008}]{Tully2008}
{Tully} R.~B.,  {Shaya} E.~J.,  {Karachentsev} I.~D.,  {Courtois} H.~M.,
  {Kocevski} D.~D.,  {Rizzi} L.,   {Peel} A.,  2008, \mn@doi [\apj]
  {10.1086/527428}, \href {http://esoads.eso.org/abs/2008ApJ...676..184T} {676,
  184}

\bibitem[\protect\citeauthoryear{{Tully} et~al.,}{{Tully}
  et~al.}{2013}]{Tully2013}
{Tully} R.~B.,  et~al., 2013, \mn@doi [\aj] {10.1088/0004-6256/146/4/86}, \href
  {http://ukads.nottingham.ac.uk/abs/2013AJ....146...86T} {146, 86}

\bibitem[\protect\citeauthoryear{{Tully}, {Courtois}, {Hoffman}  \&
  {Pomar{\`e}de}}{{Tully} et~al.}{2014}]{Laniakea}
{Tully} R.~B.,  {Courtois} H.,  {Hoffman} Y.,   {Pomar{\`e}de} D.,  2014,
  \mn@doi [\nat] {10.1038/nature13674}, \href
  {http://adsabs.harvard.edu/abs/2014Natur.513...71T} {513, 71}

\bibitem[\protect\citeauthoryear{{Tully}, {Courtois}  \& {Sorce}}{{Tully}
  et~al.}{2016}]{Tully2016}
{Tully} R.~B.,  {Courtois} H.~M.,   {Sorce} J.~G.,  2016, preprint, \href
  {http://esoads.eso.org/abs/2016arXiv160501765T} {} (\mn@eprint {arXiv}
  {1605.01765})

\bibitem[\protect\citeauthoryear{{Uzan}}{{Uzan}}{2009}]{Uzan2009}
{Uzan} J.-P.,  2009, \mn@doi [\ssr] {10.1007/s11214-009-9503-z}, \href
  {http://adsabs.harvard.edu/abs/2009SSRv..148..249U} {148, 249}

\bibitem[\protect\citeauthoryear{{Vittorio}, {Juszkiewicz}  \&
  {Davis}}{{Vittorio} et~al.}{1986}]{Vittorio1986}
{Vittorio} N.,  {Juszkiewicz} R.,   {Davis} M.,  1986, \mn@doi [\nat]
  {10.1038/323132a0}, \href
  {http://ukads.nottingham.ac.uk/abs/1986Natur.323..132V} {323, 132}

\bibitem[\protect\citeauthoryear{{Wojtak}, {Knebe}, {Watson}, {Iliev},
  {He{\ss}}, {Rapetti}, {Yepes}  \& {Gottl{\"o}ber}}{{Wojtak}
  et~al.}{2014}]{Wojtak2014}
{Wojtak} R.,  {Knebe} A.,  {Watson} W.~A.,  {Iliev} I.~T.,  {He{\ss}} S.,
  {Rapetti} D.,  {Yepes} G.,   {Gottl{\"o}ber} S.,  2014, \mn@doi [\mnras]
  {10.1093/mnras/stt2321}, \href
  {http://esoads.eso.org/abs/2014MNRAS.438.1805W} {438, 1805}

\bibitem[\protect\citeauthoryear{{Zel'Dovich}}{{Zel'Dovich}}{1970}]{ZA}
{Zel'Dovich} Y.~B.,  1970, \aap, \href
  {http://adsabs.harvard.edu/abs/1970A%26A.....5...84Z} {5, 84}

\bibitem[\protect\citeauthoryear{{Zu}, {Weinberg}, {Jennings}, {Li}  \&
  {Wyman}}{{Zu} et~al.}{2013}]{infall_Zu2013}
{Zu} Y.,  {Weinberg} D.~H.,  {Jennings} E.,  {Li} B.,   {Wyman} M.,  2013,
  preprint, \href {http://adsabs.harvard.edu/abs/2013arXiv1310.6768Z} {}
  (\mn@eprint {arXiv} {1310.6768})

\bibitem[\protect\citeauthoryear{{van de Weygaert} \& {Bertschinger}}{{van de
  Weygaert} \& {Bertschinger}}{1996}]{vdWeygaert1996}
{van de Weygaert} R.,  {Bertschinger} E.,  1996, \mn@doi [\mnras]
  {10.1093/mnras/281.1.84}, \href
  {http://adsabs.harvard.edu/abs/1996MNRAS.281...84V} {281, 84}

\bibitem[\protect\citeauthoryear{{van de Weygaert} \& {Schaap}}{{van de
  Weygaert} \& {Schaap}}{2009}]{vs2009}
{van de Weygaert} R.,  {Schaap} W.,  2009, in {Mart{\'{\i}}nez} V.~J.,  {Saar}
  E.,  {Mart{\'{\i}}nez-Gonz{\'a}lez} E.,   {Pons-Border{\'{\i}}a} M.-J.,  eds,
   Lecture Notes in Physics, Berlin Springer Verlag Vol. 665, Data Analysis in
  Cosmology. pp 291--413, \mn@doi{10.1007/978-3-540-44767-2_11}

\makeatother
\end{thebibliography}

\bsp

\label{lastpage}

\end{document}